\pdfoutput=1
\documentclass[%
 reprint,
superscriptaddress,
 amsmath,
 amssymb,
 aps,
 showkeys,
 pra
]{revtex4-2}

\usepackage{graphicx}
\usepackage{bm}
\usepackage{amsmath}

\usepackage{hyperref}

\usepackage{gensymb}
\usepackage{soul}
\usepackage[version=4]{mhchem}
\usepackage{xcolor}

\renewcommand{\vec}{\bm}
\newcommand{\mat}{\bm}

\def\tv{{\tau_{v}}}
\def\tmopt{{\tau_{\rm m}^{\rm opt}}}
\def\tm{{\tau_{\rm m}}}
\def\tr{{\tau_{\rm r}}}
\def\tc{{\tau_{\rm c}}}
\def\tL{{\tau_{\ell}}}

\def\xT{X_{\rm T}}
\def\RT{R_{\rm T}}

\def\kf{{k_{\rm f}}}
\def\kr{{k_{\rm r}}}
\def\KD{{K_{\rm D}}}
\def\KDI{{K_{\rm D}^{\rm I}}}
\def\KDA{{K_{\rm D}^{\rm A}}}

\newcommand{\avg}[1]{\langle #1\rangle}

\newcommand{\fref}[1]{\textcolor{blue}{Fig.~\ref{fig:#1}}}

\newcommand{\flabel}[1]{\label{fig:#1}}
\newcommand{\eref}[1]{Eq.~\ref{eqn:#1}}

\newcommand{\erefstwo}[2]{Eqs.~\ref{eqn:#1}~and~\ref{eqn:#2}}

\newcommand{\elabel}[1]{\label{eqn:#1}}

\begin{document}

\title{Trade-offs between cost and information in cellular prediction}

\author{Age J. Tjalma}
\affiliation{AMOLF, Science Park 104, 1098 XG Amsterdam, The Netherlands}
\author{Vahe Galstyan}
\affiliation{AMOLF, Science Park 104, 1098 XG Amsterdam, The Netherlands}
\author{Jeroen Goedhart}
\affiliation{AMOLF, Science Park 104, 1098 XG Amsterdam, The Netherlands}
\author{Lotte Slim}
\affiliation{AMOLF, Science Park 104, 1098 XG Amsterdam, The Netherlands}
\author{Nils B. Becker}
\affiliation{Theoretical Systems Biology, German Cancer Research Center, 69120 Heidelberg, Germany}
\author{Pieter Rein ten Wolde}
\email{p.t.wolde@amolf.nl}
\affiliation{AMOLF, Science Park 104, 1098 XG Amsterdam, The Netherlands}

\date{\today}

\begin{abstract}
 Living cells can leverage correlations in environmental fluctuations to predict the future environment and mount a response ahead of time. To this end, cells need to encode the past signal into the output of the intracellular network from which the future input is predicted. Yet, storing information is costly while not all features of the past signal are equally informative on the future input signal. Here, we show, for two classes of input signals, that cellular networks can reach the fundamental bound on the predictive information as set by the information extracted from the past signal: push-pull networks can reach this information bound for Markovian signals, while networks that take a temporal derivative can reach the bound for predicting the future derivative of non-Markovian signals. However, the bits of past information that are most informative about the future signal are also prohibitively costly. As a result, the optimal system that maximizes the predictive information for a given resource cost is, in general, not at the information bound.  Applying our theory to the chemotaxis network of {\it Escherichia coli} reveals that its adaptive kernel is optimal for predicting future concentration changes over a broad range of background concentrations, and that the system has been tailored to predicting these changes in shallow gradients.
\end{abstract}

\maketitle


Single-celled organisms live in a highly dynamic
environment to which they continually have to respond and adapt. To
this end, they employ a range of response strategies, tailored to the
temporal structure of the environmental variations. When these variations are highly regular, such as the daily light
variations, it
becomes beneficial to develop a clock from which the time and hence
the current and future environment can be inferred \cite{monti_robustness_2018,Pittayakanchit:2018ib}. In the other
limit, when the fluctuations are entirely unpredictable, cells have no
choice but to resort to either the strategy of detect-and-respond or
the bet-hedging strategy of stochastic switching between different
phenotypes \cite{kussell_phenotypic_2005}. Yet arguably the most fascinating strategy lies in between these two extremes. When the environmental fluctuations happen with some
regularity, then it becomes feasible to predict the future environment
and initiate a response ahead of time. While it is commonly believed
that only higher organisms can predict the future, experiments have
vividly demonstrated that even single-cell organisms can leverage
temporal correlations in environmental fluctuations in order to
predict, e.g., future nutrient levels \cite{tagkopoulos_predictive_2008, mitchell_adaptive_2009}.

The ability to predict future signals can provide a fitness benefit \cite{bialek_biophysics_2012}. The capacity to anticipate changes in oxygen levels \cite{tagkopoulos_predictive_2008}, or the arrival of sugars or stress signals \cite{mitchell_adaptive_2009},  can increase the growth rate of single-celled organisms; modeling has revealed that prediction can enhance bacterial chemotaxis \cite{becker_optimal_2015}. Yet, a predict-and-anticipate strategy is only advantageous if the cell can reliably predict the future on timescales that are longer than the time it takes to mount a response.
What fundamentally limits the accuracy of cellular prediction remains, however, poorly understood.

While the cell needs to predict the future environment, it can only sense the present and remember the past  (\fref{overview}A).  Consequently, for a given amount of information the cell can store about the present and past signal, there is a maximum amount of information it can possibly have about the future  \cite{bialek_biophysics_2012,tishby_information_1999} (\fref{overview}C-I). 
This {\em information bound} is determined by the temporal structure of the environmental fluctuations  \cite{tishby_information_1999,sachdeva_optimal_2021}. 

How close cells can come to this bound depends on the design of the intracellular biochemical network that senses and processes the environmental signals (\fref{overview}B).  To maximize the predictive power the cell must use its memory effectively: it should extract only those characteristics from the present and 
past signal that are most informative about the future \cite{becker_optimal_2015}. Whether it can do so, is determined by the topology of the signaling network.
Moreover, like any information processing device, biochemical networks require resources to be built and run. Molecular components are needed to construct the network, space is required to accommodate the components, time is needed to process the information, and energy is required to synthesize the components and operate the network \cite{govern_optimal_2014}. These resources constrain the design and performance of any
biochemical network, and the capacity to sense and process information is no exception (\fref{overview}C-II).

Cellular signaling systems provide a unique opportunity for revealing the resource requirements for prediction. Cells live in a highly dynamic environment, with temporal statistics that are expected to vary markedly. Moreover, signaling networks have distinct topologies, which are likely tailored to the temporal statistics of the environment \cite{becker_optimal_2015}. In addition, for cellular systems we can actually quantify the information processing capacity as a function of the resources that are necessary to build and run them---protein copies,
time, and energy \cite{govern_optimal_2014,Malaguti.2021}. Cellular systems are thus ideal for elucidating the relationships between future and past information, system design (i.e. network topology) and resource constraints. Here, we derive the bound on the prediction precision as set by the information extracted from the past signal for two types of input signals. We will determine how close cellular networks can come to this bound, and how this depends on the topology of the network and the resources to build and run it.

\begin{figure*}[t]
	\centering
	\includegraphics[width=\linewidth]{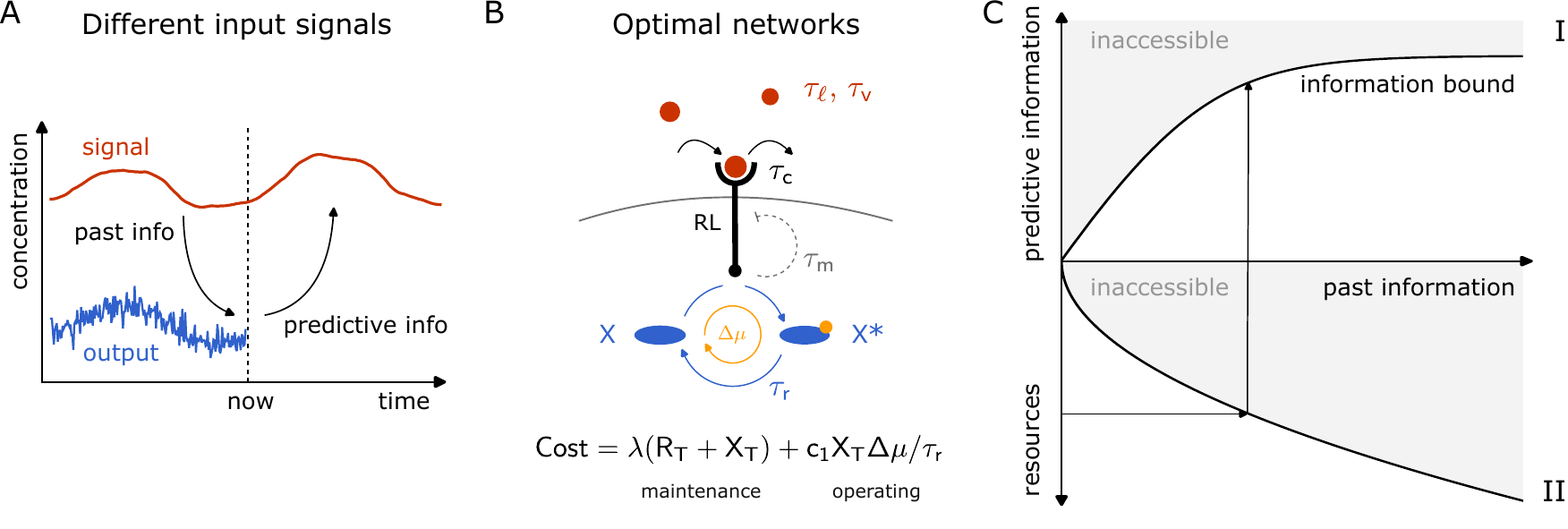}
	\caption{\textbf{Cells use biochemical networks to remember the
			past and predict the future.} (A) Cells compress the past input
		into the dynamics of the signalling network from which the future
		input is then predicted. (B) The optimal topology of the network
		for predicting the future signal depends on the temporal
		statistics of the input signal. Push-pull networks, consisting of chemical modification cycles or GTPase cycles, can optimally
		predict the future value of Markovian signals, with correlation
		time $\tL$; derivative-taking networks, like the {\it
			E. coli} chemotaxis system, can optimally predict the future
		derivative of non-Markovian signals, with correlation
		time $\tv$.  The push-pull network consists of a receptor that
		drives a downstream phosphorylation cycle. The ligand binds the
		receptor with a correlation time $\tc$. The push-pull network,
		driven by ATP turnover, integrates the receptor with an
		integration time $\tr$. The chemotaxis system is a push-pull
		network, yet augmented with negative feedback on the receptor activity
		via methylation on a timescale $\tm$, as indicated by the dashed
		grey line. The total resource cost consists of a maintenance cost
		of receptor and readout synthesis at the growth rate $\lambda$,
		and an operating cost of driving the cycle.  (C) The predictive
		information on the future signal $I_\text{pred}$ is fundamentally
		bounded by how much information $I_\text{past}$ it has about the past signal
		(panel I), which in turn is limited by the resources necessary to
		build and operate the biochemical network (panel II) \cite{bialek_biophysics_2012}.}
	\label{fig:overview}
\end{figure*}

We find that for the two classes of input signals studied, cellular
networks exists that can reach the information bound, yet reaching
the bound is exceedingly costly.  The first class of input signals
consists of Markovian signals. Using the Information Bottleneck Method (IBM)
\cite{tishby_information_1999, chechik_information_2005}, we first
show that the system that reaches the information bound
copies the most recent input signal into the output from which the
future input is predicted. Push-pull networks consisting of chemical
modification or GTPase cycles, which are ubiquitous in prokaryotic
and eukaryotic cells \cite{goldbeter_amplified_1981,
	alon_introduction_2006}, should be able to reach the information
bound, because they are at heart copying devices
\cite{govern_optimal_2014,Malaguti.2021}.  Yet, copying the most
recent input into the output is extremely costly, because the
operating cost, as set by the chemical power to drive the cycle,
diverges at high copying speed. More surprisingly, our results show that the predictive
and past information can be raised simultaneously by moving away
from the information bound, even when the operating cost is
negligible: the optimal system that maximizes the predictive
information for a given protein synthesis cost is, in general, not
at the information bound. 
The number of bits of past information per
protein cost can be raised by increasing the integration time. While
this decreases the predictive power per bit of past information,
thereby moving the system away from the information bound, it can
increase the total predictive information per protein cost. Our
analysis thus highlights that not all bits of past information are
equally costly, nor predictive.

Living cells that navigate their environment typically experience
signals with persistence as generated by their own motion, which
motivated us to study a simple class of non-Markovian signals.
Moreover, these cells can typically detect changes in the
concentration over a range of background concentrations that is
orders of magnitude larger than the change in the concentration over
the orientational correlation time of their movement. Our analysis reveals that in such a scenario the optimal kernel that allows the system to reach the information bound on predicting the future input derivative is a perfectively adaptive, derivative-taking kernel, precisely as the bacterium {\it E. coli} employs \cite{segall_temporal_1986}. We again find, however, that reaching the information bound is prohibitively costly. The reason is that taking an instantaneous derivative, which is the characteristic of the input that is most informative about the future derivative, reduces the gain to zero because the system instantly adapts; the response becomes thwarted by biochemical noise. The optimal system that maximizes the predictive information under a resource constraint thus emerges from a trade-off between taking a derivative that is recent and one that is reliable. Finally, our analysis reveals that the {\it E. coli} chemotaxis system has been optimally designed to predict future concentration changes in shallow gradients.

\section*{Results}
We focus on cellular signaling systems that respond linearly to changes in the input signal \cite{Govern:2014ef,Hinczewski:2014iq,Malaguti.2021,mattingly_escherichia_2021,Wang.2022}. These systems not only allow for analytical results, but also describe information transmission often remarkably well \cite{Tanase-Nicola2006,Ziv2007,DeRonde2010,Wang.2022}. The output of these systems can be written as
\begin{equation}
	x(t) = \int_{-\infty}^t dt^\prime k(t-t^\prime) \ell(t^\prime) + \eta_x(t),
	\elabel{linsys}
\end{equation}
where $k(t)$ is the linear response function, $\ell(t)$ the input signal,  and $\eta_x(t)$ describes the noise in the output. We will consider stationary signals with different temporal correlations, obeying Gaussian statistics.

Any  prediction about the future state of the environment must be based
on information obtained from its past
(\fref{overview}C-I). In particular, the cell needs to predict the input $\ell_\tau\equiv \ell(t+\tau)$ at a time $\tau$ into the future from the current output $x_0 \equiv x(t)$, which itself depends on the input signal in the past, $\vec{L}_p\equiv(\ell(t),\ell(t^\prime),\cdots)$, with $t> t^\prime>\cdots$. The (qualitative) shape of the integration kernel $k(t)$, e.g. exponential, adaptive or oscillatory, is determined by the topology of the signaling network \cite{becker_optimal_2015}. The kernel shape describes how the past signal is mapped onto the current output, and hence which characteristics of the past signal the cell uses to predict the future signal. To maximize the accuracy of prediction, the cell should extract those features that are most informative about the future signal. These depend on the statistics of the input signal.

Deriving the upper bound on the predictive information as set by the past information is an optimisation problem, which can be solved using the IBM \cite{tishby_information_1999}. It entails the maximization of an objective function ${\cal L}$:
\begin{equation}
	\elabel{object}
	\max_{P(x_0|\vec{L}_p)}  \left[ {\cal L} \equiv   I(x_0; \ell_\tau) - \gamma I(x_0; \vec{L}_p)\right].
\end{equation}
Here,   $I_{\rm pred} \equiv I(x_0;\ell_\tau)$ is the predictive information, which is the  mutual information between the system's current output $x_0$ and the future ligand
concentration $\ell_\tau$.  The past information $I_{\rm past}\equiv
I(x_0; \vec{L}_p)$ is the mutual information between $x_0$ and the
trajectory of past ligand concentrations $\vec{L}_p$. The Lagrange
multiplier $\gamma$ sets the relative cost of storing past 
over obtaining predictive information. Given a value of $\gamma$, the objective function in
\eref{object} is maximized by optimizing the conditional probability distribution of the output given the past input trajectory, $P(x_0|\vec{L}_p)$. For the linear systems considered here, this corresponds to optimizing the  mapping of the past input signal onto the current output via the integration kernel $k(t)$. Since our model obeys Gaussian statistics, we use the Gaussian IBM to derive the optimal kernel $k^{\rm opt}(t)$ and the  {\em information bound}, defined to be the maximum predictive information as set by the past information \cite{chechik_information_2005} (see Appendix \ref{sec:IBM}).

\subsection*{Markovian signals}
\subsubsection*{Optimal prediction of Markovian signals: biochemical copying}
Arguably the most elementary type of signal, albeit perhaps the hardest to predict, is a Markovian signal. We consider a Markovian signal $\ell(t)$, of which the deviations $\delta \ell (t)=\ell(t) - \bar{\ell}$  from its mean $\bar{\ell}$ follow an Ornstein-Uhlenbeck (OU) process:
\begin{equation}
	\elabel{OUprocess}
	\delta \dot{\ell} = - \delta \ell(t) /\tau_\ell + \eta_{\ell}(t),
\end{equation}
where $\tau_\ell$ is the correlation time of the fluctuations, and $\eta_{\ell}(t)$ is Gaussian white noise, $\avg{\eta(t)\eta(t^\prime)} = 2 \sigma^2_\ell / \tL \,\delta (t-t^\prime)$, with $\sigma^2_\ell$ the amplitude of the signal fluctuations. 
This input signal obeys Gaussian statistics, characterized by $\avg{\delta \ell(0)\delta \ell(t)}=\sigma^2_\ell  \exp(-t / \tau_\ell)$. The optimal mapping is therefore a linear one. Utilizing the Gaussian IBM framework \cite{chechik_information_2005}, we find that the optimal integration kernel is given by (see Appendix \ref{sec:IBM:markov})
\begin{equation}
	k^{\rm opt} (t-t^\prime) = a \delta (t-t^\prime).
	\elabel{koptPPN}
\end{equation}
This optimal integration kernel corresponds to a signaling system that copies the current input into the output. This is intuitive, since for a Markovian signal there is no additional information in the past signal that is not already contained in the present one. The prefactor $a$ determines the gain $\partial \bar{x} / \partial\bar{\ell}$, which together with the noise strength $\sigma^2_{\eta_x}$ (\eref{linsys}) and the signal amplitude $\sigma^2_\ell$ set the magnitude of the past and predictive information, $I_{\rm past}$ and $I_{\rm pred}$, respectively (see Appendix \ref{sec:IBM:gaussian}). 

\fref{PPN}-I shows the maximum predictive information as set by the past information. This information bound applies to any linear system that needs to predict a Markovian signal. How close can biochemical systems come to this bound?

\begin{figure}[t]
	\centering
	\includegraphics[width=\linewidth]{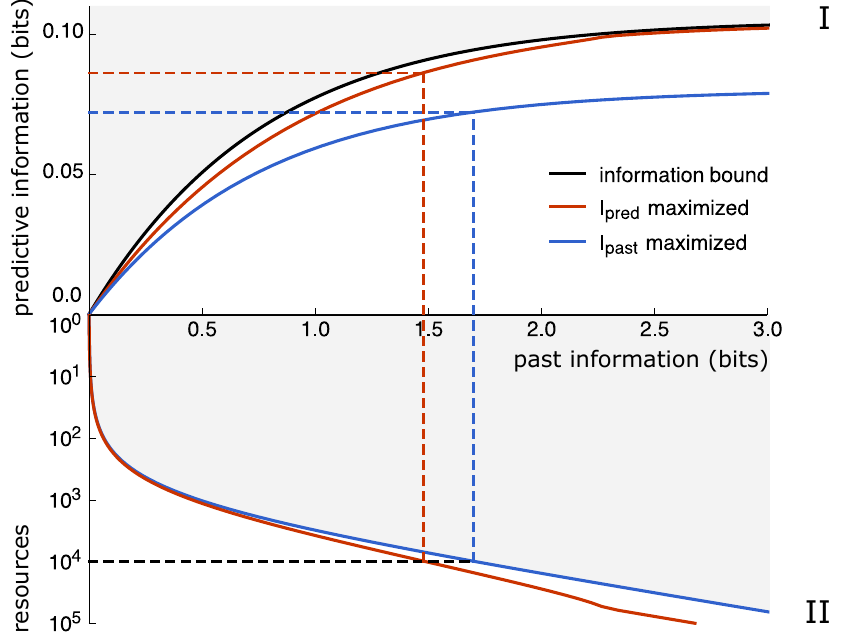}
	\caption{\textbf{The optimal push-pull network is not at the
			information bound.}  Panel I: The black line is the information
		bound that maximizes the predictive information
		$I_{\rm pred} = I(x_0;\ell_\tau)$ for a given past information
		$I_{\rm past}=I(x_0;\vec{L}_p)$. The red curve shows $I_{\rm pred}$
		against $I_{\rm past}$ for systems in which $I_{\rm pred}$ has been
		maximized for a given resource cost $C=\RT+\xT$. The blue curve shows
		$I_{\rm pred}$ versus $I_{\rm past}$ for systems where
		$I_{\rm past}$ has been maximized for a given $C$. Panel II shows
		$I_{\rm past}$ against $C$ for the corresponding systems. The forecast
			interval is $\tau=\tL$. The
		optimization parameters are the ratio $\xT/\RT$, $\tr$, $p$ and $f$
		(see Appendix \ref{sec:PPN}). Parameter values:
		$(\sigma_\ell/\bar\ell)^2 = 10^{-2}, \tc/\tL=10^{-2}$.}
	\label{fig:PPN}
\end{figure}

\subsubsection*{Push-pull network can be at the information bound, yet increase the predictive and past information by moving away from it}
Although the upper bound on the accuracy of prediction is
determined by the signal statistics, how close cells can come to
this bound depends on the topology of the cellular signaling system, and the resources devoted to building and operating it. A  network motif
that could reach the information bound for Markovian
signals is the push-pull network (\fref{PPN}),
because it is at heart a copying device: it samples the input by copying the state of the input, e.g. the ligand-binding state of a receptor or the activation state of a kinase, into the activation state of the output, e.g. phosphorylation state of the readout \cite{govern_optimal_2014,ouldridge_thermodynamics_2017,Malaguti.2021}.

We model the push-pull network in the linear-noise approximation:

\begin{align}
		\delta \dot{RL} &=  b\delta \ell(t)-\delta RL (t)/\tc + \eta_{RL}(t), \\
		\dot{\delta x^*} &=\gamma \,\delta RL(t) - \delta x^*(t)/\tr + \eta_x(t). 
\end{align}
Here,  $\delta RL$ represents the number of ligand-bound receptors and $\delta x^*$ the number of modified readout molecules, defined as deviations from their mean values; $b$ and $\gamma$ are parameters that depend on the number of receptor and readout molecules, $\RT$ and $\xT$ respectively, the fraction of ligand-bound receptors $p$ and active readout molecules $f$; $\eta_{RL}$ and $\eta_x$ are Gaussian white noise terms (see Appendix \ref{sec:PPN}). Key parameters are the correlation time of receptor-ligand binding, $\tc$, and the relaxation time of $x^\ast$, $\tr$. The latter determines for how long $x^\ast$ carries information on the ligand-binding state of the receptor and thus sets the integration time. 
The readout-modification dynamics yield an exponential integration kernel $k(t) \propto \exp(-t / \tr)$, which in the limit $\tr \to 0$ reduces to a $\delta$-function, hinting that the system may reach the information bound.

How much information cells can extract from the past signal depends on the resources devoted to building and operating the network (\fref{PPN}-II). We define the total resource cost to be:
\begin{align}
	C =  \lambda (\RT + \xT)  + c_1 \xT \Delta \mu / \tr
	\elabel{cost}
\end{align}
The first term expresses the fact that over the course of the cell cycle all components need to be duplicated, which means that they have to be synthesized at a speed that is at least the growth rate $\lambda$.
The second term describes the chemical power that is necessary to run the push-pull network \cite{govern_optimal_2014,Malaguti.2021}; it depends on the flux through the network, $\xT / \tr$, and the free-energy drop $\Delta \mu$ over a cycle, e.g. the free energy of ATP hydrolysis in the case of a phosphorylation cycle. The coefficient $c_1$ describes the relative energetic cost of synthesising the components during the cell cycle versus that of running the system. For simplicity, we first consider the scenario that the cost is dominated by that of protein synthesis, setting $c_1 \to  0$. While in this scenario $\RT + \xT$ is constrained, $\xT / \RT$ and other system parameters are free for optimization.

The available resources put a hard bound on the information $I_{\rm past}$ that can be extracted from the past signal, which in turn sets a hard limit on the predictive information $I_{\rm pred}$ (\fref{overview}C).   To maximize the predictive information, it therefore seems natural to maximize the past information $I_{\rm past}$ for a given resource cost $C$.  The blue line in \fref{PPN}-II shows the result for the push-pull network. We then compute the corresponding predictive information for the systems along this line, which is  the blue line in \fref{PPN}-I. Strikingly, the resulting information curve lies far below the information bound, i.e. the upper bound on the predictive information as set by the past information (black line, \fref{PPN}-I). This shows that systems that maximize past information under a resource constraint, do not in general also maximize predictive information. It implies that not all bits of past information are equally predictive about the future.

Precisely because not all bits of past information are equally predictive about the future,  it is paramount to directly maximize the predictive information for a given resource cost in order to obtain the most efficient prediction device. This yields the red lines in panels I and II in \fref{PPN}. It can be seen that the predictive information is higher while the past information is lower, as compared to the information curves of the systems optimized for maximizing the past information under a resource constraint (blue lines). It reflects the idea that not all bits are equally predictive. More surprisingly, while the bound on the predictive information as set by the resource cost (red line panel I) is close to the bound on the predictive information as set by the past information (black line), it does remain lower. This is surprising, because the push-pull network is a copying device \cite{govern_optimal_2014, ouldridge_thermodynamics_2017}, which can, as we will also show below,  reach the latter bound. These two observations together imply that not all bits of past information are equally costly. If they were, the cell would select under the two constraints the same bits based on their predictive information content, and the bound on the predictive information as set by the resource cost would overlap with that as set by the past information.

We thus find that not all bits of past information are equally
predictive, nor equally costly. As we show next, it implies that the optimal information processing system faces a trade-off between using those bits of past information that are most informative about the future and those that are cheapest.

\begin{figure*}[tbhp]
	\centering
	\includegraphics[width=\linewidth]{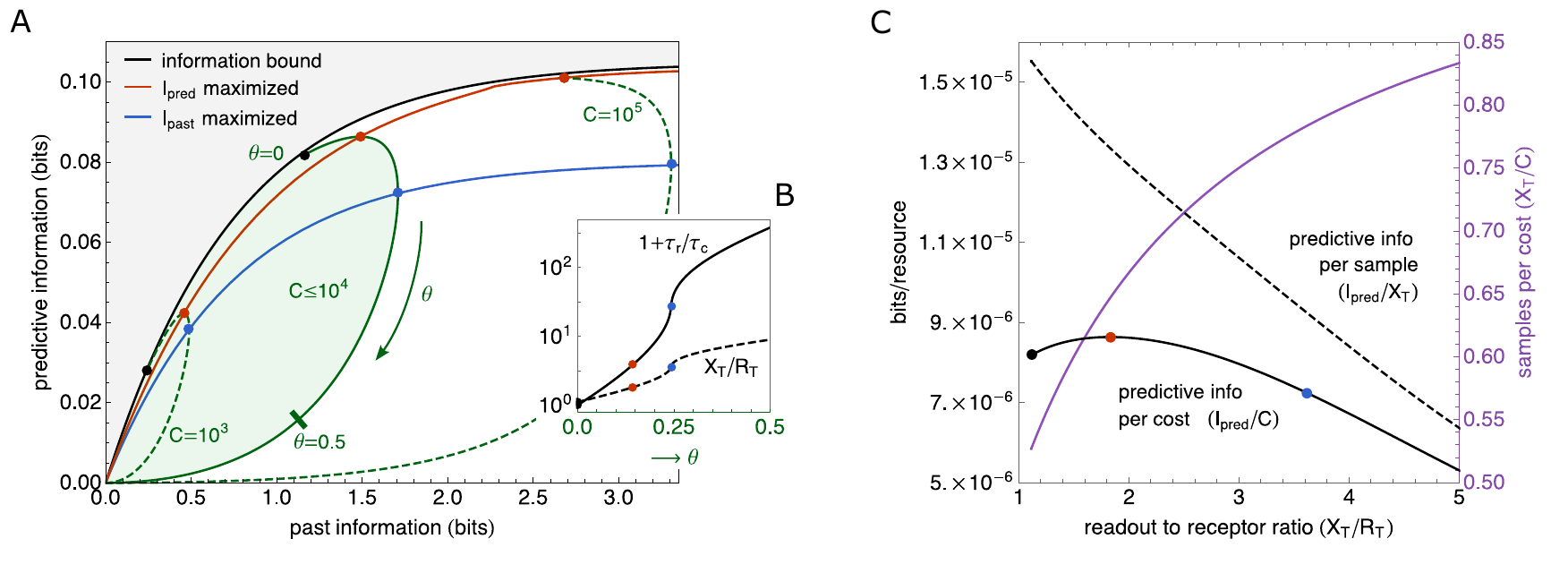}
	\caption{\textbf{The push-pull network maximizes the predictive power under a resource constraint by moving away from the information bound.} (A)
		The region of accessible predictive information
		$I_{\rm pred}=I(x_0;\ell_\tau)$ and past information
		$I_{\rm past}=I(x_0;\vec{L}_p)$ in the push-pull network under a
		resource constraint $C\leq(\RT + \xT)$, for the Markovian signals
		specified by \eref{OUprocess} (green). The black line is the
		information bound at which $I_{\rm pred}$ is maximized for a given
		$I_{\rm past}$. The push-pull network can be at the information
		bound (black points), but maximizing $I_{\rm pred}$ for a resource
		constraint $C$ moves the system away from it. The red and blue lines
		connect, respectively, the points where $I_{\rm pred}$ and
		$I_{\rm past}$ are maximized along the green isocost lines (the
		contourlines of constant $C$); they correspond to the red and blue
		lines in \fref{PPN}, respectively.  The accessible region of
		$I_{\rm pred}$ and $I_{\rm past}$ for a given $C$ has been obtained
		by optimizing over $\tr$, $p$, $f$, and $\xT/\RT$. The forecast interval is $\tau=\tL$. (B) The
		integration time $\tr$ over the receptor correlation time $\tc$,
		$\tr / \tc$, and the ratio of the number of readout and receptor
		molecules, $\xT / \RT$, as a function of the distance $\theta$ along
		the isocost line corresponding to $C=10^4$ in panel A; the red and
		blue points denote where $I_{\rm pred}$ and $I_{\rm past}$ are
		maximized along the contourline, respectively. For $\theta \to 0$,
		$\tr \to 0$: the system is an instantaneous responder, which is
		essentially at the information boundary; as predicted by the optimal
		resource allocation principle, $\xT = \RT$. The
		system can increase $I_{\rm pred}$ and $I_{\rm past}$ by increasing
		$\tr$ and $\xT/\RT$. (C) While this decreases the predictive
		information $I_{\rm pred}$ per physical bit of past information,
		$I_{\rm pred}/\xT$ (dashed line), increasing $\xT / \RT$ does
		increase the number of physical bits per resource cost, $\xT / C$
		(purple line). This trade-off gives rise to an optimal
		predictive information per resource cost, $I_{\rm pred}/C$
		(red dot on solid black line). 
		Parameter values unless specified:
		($\sigma_\ell/ \bar{\ell})^2=10^{-2}, \tc/\tL=10^{-2}$. }
	\label{fig:info_org}
\end{figure*}

\subsubsection*{Trade-off between cost and predictive power per bit}
To understand the connection between predictive and past information, and resource cost, we map out the region in the information plane that
can be reached given a resource constraint $C$
(\fref{info_org}A, green region). We immediately make two
observations. Firstly, the system can indeed reach the
information bound. Secondly, the system can increase both the
past and the predictive information by moving away from the bound.
To elucidate these two observations, we investigate the system along
the isocost line of $C=10^4$, which together with the information
bound envelopes the accessible region for the maximum
resource cost $C\leq 10^4$.

Along the isocost line, the ratio of the number of readout over receptor molecules is $\xT / \RT = 2 \sqrt{p/(1-p)}\sqrt{1+\tr/\tc}$ (see Appendix \ref{sec:PPN:optallocation}). This can be understood intuitively using
the optimal resource allocation principle
\cite{govern_optimal_2014}. It states that in a sensing system
that employs its proteins optimally, the total number of
independent concentration measurements at the level of the receptor during the integration time $\tr$,
$\RT (1+\tr/\tc)$, equals the number of readout molecules $\xT$ that
store these measurements, so that neither the receptors nor the
readout molecules are in excess. This
design principle specifies, for a given integration time $\tr$, the
ratio 
$\xT/\RT$ at which the readout molecules sample
each receptor molecule roughly once every receptor correlation time
$\tc$.

While the optimal allocation principle gives the optimal ratio
$\xT/\RT$ of the number of readouts over receptors for a {\em
	given} integration time $\tr$, it does not prescribe what the
optimal integration time  $\tr^{\rm opt}$, and hence  (globally) optimal
ratio $\xT^{\rm opt} / \RT^{\rm opt}$, is  that maximizes
$I_\text{pred}$ for a given resource constraint
$C=\RT+\xT$. \fref{info_org}B shows that as the distance $\theta$ along the isocost line is increased, $\tr$ and hence $\xT / \RT$ increase monotonically. 
Near the information
bound, corresponding to $\theta = 0$, the integration time $\tr$
is zero and the number of readout molecules equals the number of
receptor molecules: $\xT = \RT$. In this limit, the push-pull network
is an instantaneous responder, with an integration kernel given by \eref{koptPPN}; only the
finite receptor correlation time $\tc$ prevents the system from
fully reaching the information bound. Yet, as $\theta$ increases and the system moves away from the bound, the predictive and past information first rise along the contour, and thus with $\xT/\RT$ and $\tr$, before they eventually both fall.

To understand why the predictive and past information first rise and then fall with $\xT/\RT$ and $\tr$, we note that each readout molecule constitutes 1
physical bit and that its binary state (phosphorylated or not)
encodes at most 1 bit of information on the ligand
concentration. The number of readout molecules $\xT$ thus sets a hard
upper bound on the sensing precision and hence the predictive
information. To raise this bound, $\xT$ must be increased. For a given
resource constraint $C=\RT+\xT$, $\xT$ can only be increased if the
number of receptors $\RT$ is simultaneously decreased. However, the
cell infers the concentration not from the readout molecules directly,
but via the receptor molecules: a readout molecule is a sample of
the receptor that provides at most 1 bit of information about the
ligand-binding state of a receptor molecule, which in turn provides at
most 1 bit of information about the input signal. To raise the lower
bound on the predictive information, the information on the input must
increase at both the receptor and the readout level.

To elucidate how this can be achieved, we note that the maximum number
of independent receptor samples and hence concentration measurements
is given by $N_{\rm I}^{\rm max} = {\rm min}(\xT,\RT (1+\tr/\tc))$
\cite{govern_optimal_2014}. For $\theta > 0$, the system can
increase $N_{\rm I}^{\rm max}$ if, and only if, $\xT$ and
$\RT (1+\tr / \tc)$ can be raised simultaneously. This can be
achieved, while obeying the constraint $C = \xT + \RT$, by decreasing
$\RT$ yet increasing $\tr$ (\fref{info_org}B). This is the mechanism
of time averaging, which makes it possible to increase the number of
independent receptor samples \cite{Malaguti.2021}, and explains why
both the predictive and the past information initially increase
(\fref{info_org}C). However, as $\tr$ is raised further, the
receptor samples become older: the readout molecules increasingly
reflect receptor states in the past that are less informative about
the future ligand concentration. The collected bits of past information have
become less predictive about the future (\fref{info_org}C). For
a given resource cost, the cell thus faces a trade-off between
maximizing the number of physical bits of past information
(i.e. the receptor samples $\xT$) and the predictive information per
bit.  This antagonism gives rise to an optimal integration time
$\tr^{\rm opt}$ that maximizes the total predictive information
$I_{\rm pred}$ (\fref{info_org}C).

Interestingly, while $I_{\rm pred}$ decreases beyond $\tr^{\rm opt}$,
the past information $I_{\rm past}$ first continues to rise because
$N_{\rm I}^{\rm max}$ still increases. However, when the integration
time becomes longer than the input signal correlation time, the
correlation between input and output will be lost and $I_{\rm past}$
will fall too.

\subsubsection*{Chemical power prevents the system from reaching the information bound} So far, we have only considered the cost of maintaining the cellular system, the protein cost $C=\RT+\xT$. Yet, running a push-pull network also requires energy. As \eref{cost} shows, the running cost scales with the flux around the phosphorylation cycle, which is proportional to the inverse of the integration time, $\tr^{-1}$. The power thus diverges for $\tr\to 0$. Since the information bound is reached precisely in this limit, it is clear that the chemical power prevents the push-pull network from reaching the bound (see \fref{opcost} in the appendix).

\subsection*{Non-Markovian signals}
\subsubsection*{Predicting the future change}
The push-pull network can optimally predict Markovian signals, yet not
all signals are expected to be Markovian. Especially organisms that
navigate through an environment with directional persistence will
sense a non-Markovian signal, as generated by their own
motion. Moreover, when these organisms need to climb a
concentration gradient, as {\it E. coli} during chemotaxis, then
knowing the change in the concentration is arguably more useful than
knowing the concentration itself. Indeed, it is well known that the
kernel of the {\it E. coli} chemotaxis system detects the (relative)
change in the ligand concentration by taking a temporal derivative of
the concentration \cite{segall_temporal_1986}.  However, as we will
show here, the converse statement is more subtle. If the system needs
to predict the (future) change in the signal, then the optimal kernel
is not necessarily one that is based on the derivative only: in
general, the optimal kernel uses a combination of the signal value and
its derivative. 
However, the {\it E. coli} chemotaxis system can respond to concentrations that vary between the dissociation constants of the inactive and active state of the receptors, which differ by several orders of magnitude \cite{keymer_chemosensing_2006}. This range of possible background concentrations is much larger than the typical concentration change over the orientational correlation time of the bacterium. As our analysis below reveals, in this regime the optimal kernel is a perfectly adaptive, derivative-taking kernel that is insensitive to the current signal value, precisely like that of the {\it E. coli} chemotaxis system \cite{segall_temporal_1986,1997.Barkai,2006.Endres,2012.Lan,2015.Sartori}. Our analysis thus predicts that this system has an adaptive kernel, because this is the optimal kernel for predicting concentration derivatives over a broad range of background concentrations.

To reveal the signal characteristics that control the shape of the
optimal integration kernel, we will consider the family of signals
that are generated by a harmonic oscillator:
\begin{align}
	\delta \dot{\ell} & = v(t), \elabel{NMc}\\
	\dot{v} &= - \omega_0^2 \delta \ell(t) - v (t) / \tau_{v} + \eta_v (t), \elabel{NMvc}
\end{align}
where $\delta \ell$ is the deviation of ligand concentration from its mean $\bar{\ell}$, $v$ its derivative,
$\tau_{v}$ a relaxation time, $\eta_v$ a Gaussian white noise
term, and the frequency $\omega_0^2=\sigma^2_{v}/\sigma^2_\ell$
controls the variance $\sigma^2_\ell$ of the concentration and that of
its derivative $\sigma^2_{v}$.

Using the IBM framework it can be shown that the optimal encoding
that allows the system to reach the information bound, is based on a
linear combination of the current concentration $\ell(t)$ and its
derivative $v(t)$, such that the output $x(t)$ is given by (Appendix \ref{sec:IBM:nonmarkov}):
\begin{align}
	x(t)=a \frac{\delta \ell(t)}{\sigma_\ell} + b \frac{v(t)}{\sigma_{v}}+ \eta_x(t).
	\elabel{xNM}
\end{align}
This can be understood by noting that while the signal of
\erefstwo{NMc}{NMvc} is non-Markovian in the space of $\ell$, it is
Markovian in $\ell$ and $v$: all the information on the
future signal is thus contained in the current concentration and its
derivative. To maximize the predictive
information $I_{\rm pred}=I(x_0;v_{\tau})$ between the current
output $x_0$ and the future derivative of the input $v_{\tau}$ for a
given amount of past information $I_{\rm past}=I(x_0;{\vec{L}}_p)$,
i.e to reach the information bound for predicting the future signal
derivative, the coefficients must obey 
\begin{align}
	a^{\rm opt} &= G\frac{\avg{\delta \ell(0) \delta v (\tau)}}{\sigma_\ell\sigma_{v}}\equiv G \rho_{\ell_0v_\tau}, \elabel{aopt}\\
	b^{\rm opt}&=G \frac{\avg{\delta v (0) \delta v(\tau)}}{\sigma^2_{v}}\equiv G \rho_{v_0 v_\tau}.\elabel{bopt}
\end{align}
Here, $G$ is the gain, which together with the noise $\sigma^2_{\eta_x}$ sets the scale of $I_{\rm pred}$ and $I_{\rm past}$, 
$\rho_{\ell_0v_\tau}$ is the cross-correlation coefficient
between the current concentration value $\ell_0$ and the future concentration derivative $v_\tau$ and $\rho_{v_0 v_\tau}$ that between the current and future derivative (Appendix \ref{sec:IBM:nonmarkov}).  These expressions can be understood
intuitively: if the future signal derivative that needs to be
predicted is correlated with the current signal derivative, it is
useful to include in the prediction strategy the current signal
derivative, leading to a non-zero value of $b^{\rm
	opt}$. Perhaps more surprisingly, if the future signal derivative is
also correlated with the current signal {\em value}, then the system
can enhance the prediction accuracy by also including the current
signal value, yielding a non-zero $a^{\rm opt}$. Clearly, in
general, to optimally predict the future signal change, the system
should base its prediction on both the current signal value and its
derivative.

The degree to which the systems bases its prediction on the current
value versus the current derivative depends on the relative magnitudes
of $a^{\rm opt}$ and $b^{\rm opt}$, respectively. In
	Appendix \ref{sec:signals:nonmarkov}, we show that when the concentration change over the timescale
$\tv$, $\sigma_v \tv$, is much smaller than the range of
possible concentrations $\sigma_\ell$ that the bacterium can experience,
i.e. when $\sigma_{v} \tau_{v} \ll \sigma_\ell$ such that
$\omega_0 \ll \tau_{v}^{-1}$, the cross-correlation coefficient
$\rho_{\ell_0v_\tau}$ vanishes, such that $a^{\rm opt}$ becomes
zero (see \eref{aopt}). The optimal kernel has become a perfectly
adaptive, derivative-taking kernel. We emphasize that while we have
derived this result for the class of signals defined by
\erefstwo{NMc}{NMvc}, the idea is far more generic. In particular,
while we do not know the temporal structure of the ligand statistics
that {\it E. coli} experiences, we do know that it can detect
concentration changes over a range of background concentrations that
is much wider that the typical concentration change over a run, such
that the correlation between the concentration value and its future
change is likely to be very small. As our analysis shows, a
perfectively adaptive kernel then emerges naturally from the
requirement to predict the future concentration change.

While the class of signals specified by \erefstwo{NMc}{NMvc} is
arguably limited, it does describe the biologically important regime
of chemotaxis in shallow gradients.  In the limit that
$\omega_0 \ll \tv^{-1}$, \eref{NMvc} reduces to
$\dot{v} = -v / \tau_{v}+\eta_v$. In shallow gradients, the
stimulus only weakly affects the swimming behavior, such that the
perceived signal is mostly determined by the intrinsic orientational
dynamics of the bacterium in the absence of a gradient. In this
regime, the temporal statistics of the concentration derivative $v$
is completely determined by the steepness of the concentration
gradient $g$ and the swimming statistics of the bacterium in the
absence of a gradient:
\begin{align}
	\avg{\delta v (0) \delta v(\tau)}= g^2 \bar{\ell}^2\avg{\delta v_x (0) \delta v_x(\tau)}\simeq \sigma^2_{v_x} e^{-\tau / \tau_{v_x}},
\end{align}
where the latter is the autocorrelation
function of the (positional) velocity of the bacterium in the absence of a gradient. It is a
characteristic of the bacterium, not of the environment, and has been
measured to decay exponentially with a correlation time $\tau_{v_x}$
\cite{mattingly_escherichia_2021}, precisely as our model, with
$\tv=\tau_{v_x}$, predicts. This correlation time is on the order
of the typical run time of the bacterium in the absence of a
gradient, $\tv \sim 0.9{\rm s}$
\cite{mattingly_escherichia_2021}.

\subsubsection*{Finite resources prevent the chemotaxis system from taking an instantaneous derivative and reaching the information bound}
The above analysis indicates that the chemotaxis system seems ideally
designed to predict the future concentration change, because its
integration kernel is nearly perfectly adaptive
\cite{segall_temporal_1986,1997.Barkai,2006.Endres,2012.Lan,2015.Sartori}.
But how close can this system come to the information bound for the
non-Markovian signals specified by \erefstwo{NMc}{NMvc}?

To address this, we consider a
molecular model that can accurately describe the
response of the chemotaxis system to a wide range of time-varying
signals \cite{shimizu_modular_2010, tu_modeling_2008,
	tostevin_mutual_2009,Reinhardt.2022}.  In this model, the receptors
are partitioned into clusters. Each cluster is described via a
Monod-Wyman-Changeux model \cite{monod_nature_1965}. While each
receptor can switch between an active and an inactive conformational
state, the energetic cost of having different conformations in the
same cluster is prohibitively large. Each cluster is thus either
active or inactive. Ligand binding favors the inactive state while
methylation does the opposite. Lastly, active receptor clusters can
via the associated kinase CheA phosphorylate the downstream messenger
protein CheY.

Linearizing around the steady state, we obtain:
\begin{align}
	\delta a_i(t) &= \alpha \delta m_i(t) - \beta \delta \ell(t), \elabel{da}\\
	\delta \dot{m}_i &= - \delta a_i(t) / (\alpha\tm) + \eta_{m_i} (t), \elabel{dm}\\
	\delta \dot{x}^\ast &= \gamma \sum_{i=1}^{\RT} \delta a_i(t) - \delta x^\ast(t)/ \tr+ \eta_x (t).
\end{align}
Here, $\delta a_i(t)$ and $\delta m_i(t)$ are the deviations of the
activity and methylation level of receptor cluster $i$ from their
steady-state values, and $\RT$ is the total number of receptor
clusters; $\delta \ell(t)$ and $\delta x^\ast(t)$ are, respectively,
the deviations of the ligand and ${\rm CheY_p}$ concentration from
their steady-state values; $\tm$ and $\tr$ are the timescales of
receptor methylation and ${\rm CheY_p}$ dephosphorylation; $\eta_{m_i}$
and $\eta_x$ are independent Gaussian white noise sources.  In
\eref{da}, we have assumed that ligand binding is much faster than the
other timescales in the system, so that it can be integrated
out. There is therefore no need to time average receptor-ligand
binding noise, which means that, in the absence of running costs, the optimal receptor integration time
$\tr$ is zero. In what follows, we set $\tr$ to the value measured
experimentally, $\tr \approx 100{\rm ms}$
\cite{sourjik_binding_2002,govern_optimal_2014}. We consider the non-Markovian signals specified by \erefstwo{NMc}{NMvc} in the physiologically relevant limit $\omega_0\to 0$, such that the optimal kernel is perfectly adaptive, like that of {\it E. coli}. For these signals, we determine the
accessible region of $I_{\rm past}$ and $I_{\rm pred}$ under a
resource constraint $C=\RT+\xT$ (see \fref{CN}) by optimizing over
the methylation time $\tm$ and the ratio of readout over receptor
molecules $\xT/\RT$. The forecast interval $\tau$ is set to $\tv$, but we emphasize that the optimal design is independent of the value of $\tau$ (see Appendix \ref{sec:CN:IpastIpred}).

\begin{figure*}[tbhp]
	\centering
	\includegraphics[width=\linewidth]{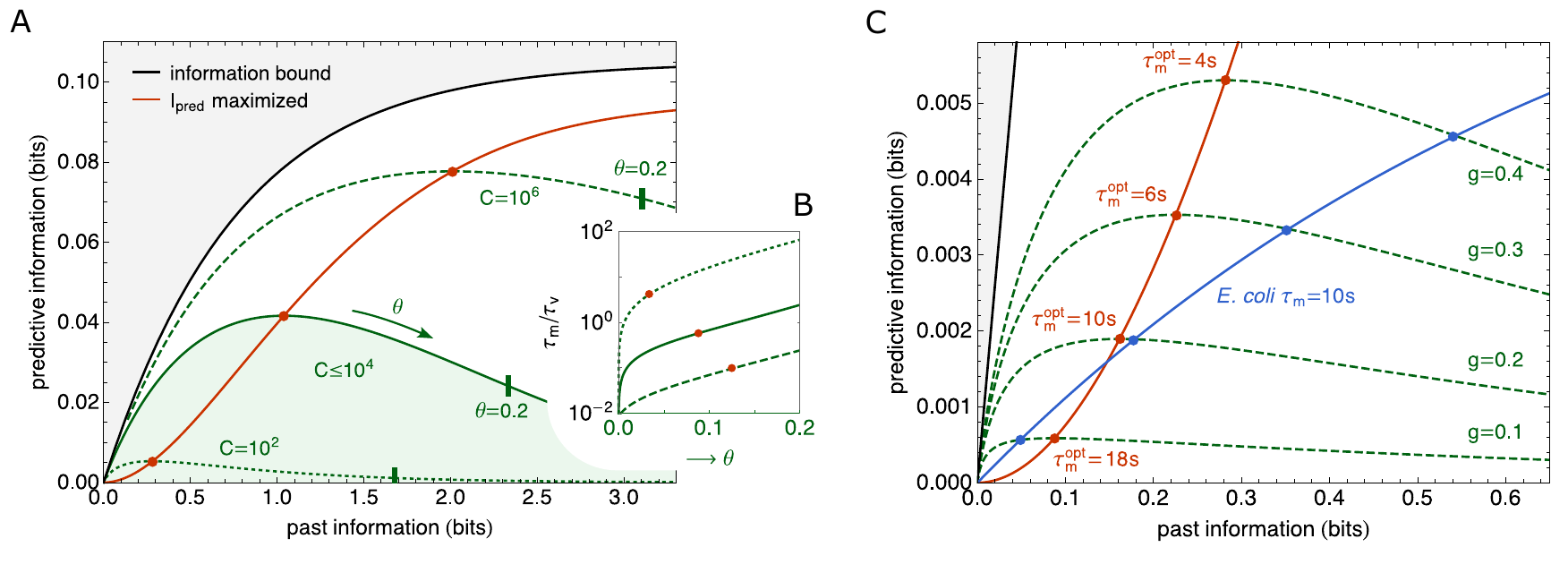}
	\caption{\textbf{Finite resources prevent chemotaxis system from reaching the information bound.} (A) The region of accessible predictive information $I_{\rm pred}=I(x_0; v_{\tau})$ and past information $I_{\rm past}=I(x;\vec{L}_p)$ for the chemotaxis system under a resource constraint $C=\RT + \xT$, for the non-Markovian signals specified by \erefstwo{NMc}{NMvc} (green). The black line shows the information bound at which $I_{\rm pred}$ is maximized for a given $I_{\rm past}$. The chemotaxis system is not at the information bound, but it does move towards it as $C$ is increased.   The red line connects the red points where $I_{\rm pred}$ is maximized for a given resource cost $C$. The accessible region of $I_{\rm pred}$ and $I_{\rm past}$ under a given resource constraint $C=\RT+\xT$ is obtained by optimizing over the methylation time $\tm$ and the ratio of readout over receptor molecules $\xT/\RT$. The forecast interval is $\tau=\tv$. (B) The methylation time $\tm$ over the input correlation time $\tv$ as a function of the distance $\theta$ along the three respective isocost lines shown in panel A. The methylation time $\tm$ increases along the isocost line, but there exists an optimal $\tm$ that maximizes the predictive information, marked by the red points; $\theta \to 0$ corresponds to the origin of panel A, $(I_{\rm pred}, I_{\rm past})=(0,0)$; the points where $\theta = 0.2$ along the isocost lines of panel A are marked with a bar. As the resource constraint is relaxed (higher $C$), the optimal $\tm$ decreases: the system moves towards the information bound, where it takes an instantaneous derivative, corresponding to $\tr, \tm\to 0$.  (C) The contourlines of $I_{\rm pred}$ and $I_{\rm past}$ for increasing values of the steepness $g$ of an exponential ligand concentration gradient $\ell(x) = \ell_0 e^{g x}$, keeping the total resource cost fixed at $C=\RT + \xT =10^4$; $\tm$ and $\xT/\RT$ have been optimized. It is seen that the maximal predictive information $I_{\rm pred}$ under the resource constraint $C$ (marked by the red points) increases with the gradient steepness. 
		The blue line shows $I_{\rm pred}$ and $I_{\rm past}$ for the {\it E. coli} chemotaxis system with $\tm=10{\rm s}$ and $\xT = \RT = 5000$ fixed at their measured values \cite{li_cellular_2004}.  Our analysis predicts that this system has been optimized to detect shallow gradients. 
		Parameter values unless specified: $\tr = 100{\rm ms}$~\cite{sourjik_binding_2002,govern_optimal_2014}; $\tv = 0.9{\rm s}$ and $\sigma^2_{v}= g^2 \bar{\ell}^2 \sigma^2_{v_x}$, with $\bar{\ell}=100\mu{\rm M}$ and $\sigma^2_{v_x}=157.1 \mu {\rm m}^2 {\rm s}^{-2}$ \cite{mattingly_escherichia_2021}; $\omega_0 \to 0$; $g$ is given in units of ${\rm mm}^{-1}$; in A, $g=4/\text{mm}.$ }
	\flabel{CN}
\end{figure*}

\fref{CN}A shows that the chemotaxis system is, in general, not at the
information bound that maximizes the predictive information  $I_{\rm pred}=I(x_0;v_{\tau})$ for a given past information $I_{\rm past}=I(x_0;\vec{L}_p)$.
The optimal systems that maximize $I_{\rm pred}$ under a
resource constraint $C$, marked by the red dots, are indeed markedly away from
the information bound. Yet, as the resource constraint is
relaxed and $C$ is increased, the optimal system moves towards
the bound. Panel B shows that the methylation time $\tm$ rises along the three respective isocost lines of panel A. It highlights that there exists an
optimal methylation time $\tmopt$ that maximizes the predictive
information $I_{\rm pred}$. Moreover, $\tmopt$ decreases as the resource constraint is relaxed. Along the respective isocost lines, $\xT/\RT$ varies only mildly (see \fref{XtRtovertheta} in the appendix).

These observations can be understood by noting that the system faces a
trade-off between taking a derivative that is recent versus one that
is robust. All the information on the future derivative, which the
cell aims to predict, is contained in the current derivative of the
signal; measuring the current derivative would allow the system to
reach the information bound. However, computing the recent derivative
is extremely costly. The cell takes the temporal derivative of the
ligand concentration at the level of the receptor via two antagonistic
reactions that occur on two distinct timescales: ligand binding
rapidly deactivates the receptor, while methylation slowly reactivates
it  \cite{tu_modeling_2008}. The receptor ligand-occupancy thus encodes the current
concentration, the methylation level stores the average concentration
over the past $\tm$, and the receptor activity reflects the difference
between the two---the temporal derivative of the signal over the timescale $\tm$. To obtain an instantaneous derivative, $\tm$ must go to zero. However, this
dramatically reduces the gain; in fact, in this limit, the gain is
zero, because the receptor activity instantly adapts to the change in
the ligand concentration. Since the push-pull network downstream of
the receptor is a device that samples the receptor 
stochastically \cite{govern_optimal_2014, malaguti_theory_2021}, the
gain, i.e. the change in the receptor activity due to the
signal, must be raised to lift the signal above the sampling noise. This requires a finite methylation time $\tm$: as we show in {Appendix \ref{sec:CN:stats}}, the gain increases monotonically with $\tm$. The trade-off between a recent derivative and a reliable one gives rise to an optimal methylation time $\tmopt$ that maximizes the predictive information for a given resource cost.

The same analysis also explains why the optimal methylation time
$\tmopt$ decreases and the predictive information increases when the
resource constraint is relaxed. The sampling noise in estimating the
average receptor activity decreases as the number of readout molecules
increases \cite{govern_optimal_2014, malaguti_theory_2021}. A smaller
gain is thus required to lift the signal above the sampling noise. In
addition, a larger number of receptors decreases the noise in the
methylation level, which also allows for a smaller gain, and hence a
smaller methylation time. These two effects together explain why
$\tmopt$ decreases and $I_{\rm pred}$ increases with $C$.

\fref{CN}A also shows that the past information
$I_{\rm past}=I(x_0;\vec{L}_p)$ does not return to zero along the
contourline of constant resource cost. Along the contourline, the
methylation time $\tm$ rises (\fref{CN}B). While the predictive
information $I_{\rm pred}$ exhibits an optimal methylation time
$\tm^{\rm opt}$, the past information $I_{\rm past}$ continues to rise
with $\tm$ because the system increasingly becomes a copying device,
rather than one that takes a temporal derivative.

\subsubsection*{Comparison with experiment} To test our theory, we study the predictive
power of the {\it E. coli} chemotaxis system as a function of the steepness of the ligand
concentration gradient, keeping 
the resource constraint at the biologically relevant value of
$C=\RT+\xT=10^4$ \cite{li_cellular_2004}. Panel C of \fref{CN} shows $I_{\rm pred}$ and
$I_{\rm past}$ for cells swimming in an exponential concentration
gradient $\ell(x) = \ell_0 e^{g x}$, for different values of the
gradient steepness $g$; along the green iso-steepness lines $\tm$ is
varied and $\xT/\RT$ is optimized to maximize $I_{\rm pred}$ and
$I_{\rm past}$, with the red dots marking $\tmopt$, while along
the blue line $\tm$ and $\xT$ and $\RT$ are fixed at their
experimentally measured values
\cite{li_cellular_2004,shimizu_modular_2010,tu_modeling_2008}. Clearly,
both the predictive and the past information rise as the gradient
steepness $g$ increases---a steeper concentration gradient yields a
larger change in the concentration, and thus a stronger signal.

More interestingly, in the optimal system $I_\text{pred}$ rises much faster with $I_\text{past}$ (red line) than in the {\it E. coli} system (blue line). A steeper
gradient $g$ yields a stronger input signal, which raises the signal
above the sampling noise more. This allows the optimal system to take a more recent derivative, with
a smaller $\tm$, which is more informative about the
future. In contrast, the methylation time $\tm$ of the {\it E. coli}
chemotaxis system is fixed. As \fref{CN}C shows, this value is beneficial
for detecting shallow gradients, $g\lesssim 0.2{\rm
	mm}^{-1}$. Moreover, in this regime, not only $I_\text{pred}$ but
also $I_\text{past}$ are close to the respective values for the optimal
system. For steeper gradients $I_\text{past}$ becomes
much higher in the {\it E. coli} system than in the optimal one,
even though $I_\text{pred}$ remains lower. The bacterium increasingly
collects information that is less informative about the future. Taken
together, these results strongly suggest that the system has been
optimized to predict future concentration changes in shallow gradients, which necessitate a relatively
long methylation time.

\section*{Discussion}
Cellular systems need to predict the future signal by capitalizing on information that is contained in the past signal. To this end, they need to encode the past signal into the dynamics of the intracellular biochemical network from which the future input is inferred. To maximize the predictive information for a given amount of information that is extracted, the cell should store those signal characteristics that are most informative about the future signal. For a Markovian signal obeying an Ornstein-Uhlenbeck process this is the current signal value, while for the non-Markovian signal corresponding to an underdamped particle in a harmonic well, this is the current signal value and its derivative. As we have seen here, cellular systems are able to extract these signal characteristics: the push-pull network can copy the current input into the output, while the chemotaxis network can take an instantaneous derivative. We have thus demonstrated that at least for two classes of signals, cellular systems are in principle able to extract the most predictive information, allowing them to reach the information bound.

Yet, our analysis also shows that extracting the most relevant information can be exceedingly costly. To copy the most recent input signal into the output, the integration time of the push-pull network needs to go to zero, which means that the chemical power diverges. Moreover, taking an instantaneous derivative reduces the gain to zero, such that the signal is no longer lifted above the inevitable intrinsic biochemical noise of the signalling system. In fact, taking the chemical power cost to drive the adaptation cycle into account \cite{2012.Lan,sartori_free_2015} would push the system away from the information bound even more.

While information is a resource---the cell cannot predict the future
without extracting information from the past signal---the principal
resources that have a direct cost are time, building blocks and
energy. The predictive information per protein and energy cost is
therefore most likely a more relevant fitness measure than the predictive
information per past information.  Our analysis reveals that, in
general, it is not optimal to operate at the information bound:
cells can increase the predictive information for a given resource
constraint by moving away from the bound.  Increasing the
integration time in the push-pull network reduces the chemical power
and makes it possible to take more concentration measurements per
protein copy. And increasing the methylation time in the chemotaxis
system increases the gain. Both
enable the system to extract more information from the past
signal. Yet, increasing the integration time or the methylation time
also means that the information that has been collected, is less
informative about the future signal. This interplay gives rise to an
optimal integration and methylation time, which
maximize the predictive information for a given resource
constraint. This argument also explains why the respective systems move towards the information
bound when the resource constraint is relaxed: Increasing the number of receptor and readout molecules
allows the system to take more instantaneous concentration
measurements, which makes time averaging less important, thus
reducing the integration time. Increasing the number of readout
molecules also reduces the error in sampling the receptor state. This makes it easier to detect a change in the receptor activity resulting from the signal, thus allowing for a smaller dynamical gain and a shorter methylation time.

Information theory shows that the amount of transmitted information depends not only on the characteristics of the information processing system, but also on the statistics of the input signal. While much progress has been made in characterizing cellular signalling systems, the statistics of the input signal is typically not known, with a few notable exceptions \cite{Dubuis:2013cp}. Here, we have focussed on two classes of input signals, but it seems likely that the signals encountered by natural systems are much more diverse. It will be interesting to extend our analysis to signals with a richer temporal structure \cite{sachdeva_optimal_2021}, and see whether cellular systems exist that can optimally encode these signals for prediction.

Finally, while we have analyzed the design of cellular signaling networks to optimally predict future signals, we have not addressed the utility of information for function or behavior. It is clear that many functional or behavioral tasks, like chemotaxis \cite{mattingly_escherichia_2021}, require information, but what the relevant bits of information are is poorly understood \cite{becker_optimal_2015}. Moreover, cells ultimately employ their resources---protein copies, time, and energy---for function or behavior, not for processing information per se. Here, we have shown that maximizing predictive information under a resource constraint, $C\to I_\text{past} \to I_\text{pred}$,  does not necessarily imply maximizing past information. This hints that optimizing a functional or behavioral task under a resource constraint, $C\to I_\text{pred} \to {\rm function}$,  may not imply maximizing the predictive information necessary to carry out this task.

\begin{acknowledgments}
We thank Jenny Poulton, Manuel Reinhardt, Michael Vennettilli and Daan de Groot for many useful discussions. This work is part of the
Dutch Research Council (NWO) and was performed at
the research institute AMOLF. This project has received
funding from the European Research Council (ERC)
under the European Union's Horizon 2020 research
and innovation program (grant agreement No. 885065).
\end{acknowledgments}

\newpage
\appendix
\onecolumngrid

\section{General}
\subsection{Linear signalling networks}
Since the systems studied in the main text have a single steady state, we will study them in the linear-noise approximation \cite{vanKampen1992}. For non-linear systems, the quality of the approximation improves with system size, but it can already be remarkably good for systems with only 10 copies \cite{Tanase-Nicola2006,Ziv:2007bo,DeRonde2010}. In the linear-noise approximation, we expand the rate equations to first order around the steady state of the mean-field chemical rate equations, and compute the noise at this steady state. In this approximation the network dynamics are a multidimensional Ornstein-Uhlenbeck (OU-)process:
\begin{equation}
	\elabel{signallingOU}
	\dot{\vec{\delta y}} = \mathcal G\vec{\delta s}(t) + \mathcal J \vec{\delta y}(t) + \mathcal B \vec \xi(t), 
\end{equation}
where $\vec{\delta s}(t)$ is a length $k$ vector of input signals and $\vec{\delta y}$ is the vector of all network species of length $n$, both defined in terms of deviations from their mean. The vector $\vec \xi(t)$ describes the $m$ independent white noise processes associated with the $m$ network reactions; they have zero mean, unit variance, and are delta correlated: $\avg{\xi_i (t)}=0$, $\langle \xi_i(t) \xi_j(t')\rangle = \delta_{ij}\delta(t-t')$, with $\delta_{ij}$ the Kronecker delta. The $n \times n$ matrix $\mathcal J$ is the Jacobian of the network, the $n\times k$ signal gain matrix  
$\mathcal G$ describes the strength by which each signal impacts each species directly, the $n\times m$ matrix $\mathcal B$ contains the noise strengths. The eigenvalues of the Jacobian $\mathcal J$ must be negative for the system to be stable, and we require all signals to be stationary.

\subsection{Integration kernels, power spectra, and correlation functions}
We continue by deriving the stationary auto-correlation matrix of a multidimensional OU-process, such as \eref{signallingOU}, via the networks' power spectra. The power spectrum of a real-valued random process $X(t)$ is the squared modulus of its Fourier transform: $S_x(\omega) = \langle \delta \tilde x(-\omega) \delta \tilde x(\omega) \rangle$ and $S_{x\to y}(\omega)  =\langle \delta \tilde x(-\omega) \delta \tilde y(\omega) \rangle$. Throughout this work we use the following conventions for the Fourier transform and its inverse: $\mathcal{F}\{f(t)\} \equiv \tilde{f}(\omega) = \int^\infty_{-\infty} dt f(t) \exp(- i \omega t)$ and $\mathcal{F}^{-1}\{\tilde f(\omega)\}=1/(2\pi)\int^\infty_{-\infty} d\omega \tilde{f}(\omega) \exp(i \omega t)=f(t)$. To obtain the correlation functions from the power spectra we invoke the Wiener-Khinchin theorem.

The general solution to \eref{signallingOU} is
\begin{equation}
	\elabel{timesol}
	\vec{\delta y}(t) = \int_{-\infty}^t dt' \,e^{\mathcal J (t-t')} \left(\mathcal G \vec{\delta s}(t') + \mathcal B \vec \xi(t')\right),
\end{equation}
which shows the two contributions to the time dependent solution: that of the external signal and that of the internal noise. The $n \times k$ matrix $e^{\mathcal J (t-t')} \mathcal G$ contains the integration kernels, its $(i,\,j)^\text{th}$ entry determines how the $j^\text{th}$ signal affects the $i^\text{th}$ system component over time. The $n \times m$ matrix $e^{\mathcal J (t-t')} \mathcal B$ is similar, but contains the functions that map the noise terms onto the system components. These matrices can be obtained by taking the Fourier transform of \eref{signallingOU} and solving for $\vec{\delta \tilde y}(\omega)$
\begin{align}
	i\omega \vec{\delta \tilde y}(\omega) = \mathcal{G}\vec{\delta \tilde s}(\omega) + \mathcal J \vec{\delta \tilde y}(\omega) + \mathcal{B} \vec{\tilde \xi}(\omega), \\
	\elabel{fouriersol}
	\vec{\delta \tilde y}(\omega) = (i\omega \mathbb{I}_n -\mathcal J)^{-1}\left(\mathcal{G}\vec{\delta \tilde s}(\omega) + \mathcal{B} \vec{\tilde \xi}(\omega)\right).
\end{align}
Using the convolution theorem to take the Fourier transform of \eref{timesol}, and comparing the result to \eref{fouriersol}, now shows that $\mathcal F\{e^{\mathcal J (t-t')}\}=(i\omega \mathbb{I}_n -\mathcal J)^{-1}$. We obtain for the power-spectra of the network components
\begin{equation}
	\elabel{powspecgeneral}
	\begin{split}
		\mathcal{S}_y(\omega)&=\langle \delta \tilde{\vec y}(-\omega)\delta \tilde{\vec y}(\omega)^T\rangle, \\
		&= \mathbb{G}(-\omega) \mathcal{S}_s(\omega) \mathbb{G}(\omega)^T+|\mathbb{N}(\omega)|^2,
	\end{split}
\end{equation}
with the matrices of frequency dependent gains $\mathbb G(\omega) \equiv (i\omega \mathbb{I}_n -\mathcal J)^{-1} \mathcal{G}$, and frequency dependent noise $\mathbb{N}(\omega)\equiv(i\omega \mathbb{I}_n -\mathcal J)^{-1} \mathcal{B}$. The cross terms vanish because the fluctuations of the external signal are uncorrelated from the internal noise. Furthermore, the power spectrum of a white noise process is constant, and all the noise terms are independent of one another, such that the spectral density of the noise vector is the identity matrix $\langle \tilde{\vec \xi}(-\omega) \tilde{\vec \xi}(\omega)^T\rangle =\mathbb{I}_m$. We also need to consider the cross-spectra between the signals and the network components, specifically we will need the spectra from the network to the signals 
\begin{equation}
	\elabel{crosspowspec}
	\begin{split}
		\mathcal{S}_{y \to s}(\omega)&=\langle \delta \tilde{\vec y}(-\omega)\delta \tilde{\vec s}(\omega)^T\rangle, \\
		&= \mathbb{G}(-\omega) \mathcal{S}_s(\omega).
	\end{split}
\end{equation}
From \eref{powspecgeneral} and \eref{crosspowspec} we can obtain all necessary correlation functions and (co-)variances, by taking the inverse Fourier transform of the component of interest (for a variance we can directly set $t=0$). The advantage of using this form, is that the contribution of each signal and of the noise terms appear separately. When we are for example interested in a variance that is only caused by noise, we can omit the terms depending on the signal power spectra, and vice versa. Moreover, the power spectra are usually simpler in form than the corresponding correlation functions.

The covariance and auto-correlation matrices can also be found by solving \eref{timesol} directly in the time domain; the solutions are shown here for completeness. For a derivation, see for example the work by Vennettilli et al. \cite{vennettilli_precision_2021}. In this case it is most convenient to include the signals as system components, we thus have a new Jacobian $\mathcal J'$ and a new noise strength matrix $\mathcal{B}'$ which include all network components and the signals themselves. The covariance matrix $\mathcal C$ is then obtained by solving the Lyapunov equation
\begin{equation}
	\elabel{lyapunov}
	\mathcal J' \mathcal C + \mathcal C \mathcal J'^T + \mathcal{B}' \mathcal{B}'^T = \mat 0,
\end{equation}and the correlation matrix is given by
\begin{equation}
	\elabel{autocor}
	\mathcal{C}(\tau) = e^{\mathcal{J}'\tau} \mathcal{C} \qquad \text{for }\tau>0.
\end{equation}

\section{Signals and statistics}
\subsection{Markovian signal}
For the Markovian ligand concentration dynamics we use a 1-dimensional OU-process
\begin{equation}
	\elabel{Markovsig}
	\delta \dot \ell = - \delta \ell /\tau_\ell + \eta_\ell(t),
\end{equation}
where the ligand concentration is defined in terms of the deviation from its mean $\delta \ell = \ell(t)-\bar \ell$. The correlation time is give by $\tau_\ell$, and the noise  $\eta_\ell(t)$ is derived from a unit white noise process  $\eta_\ell(t) \equiv \sigma_\ell\sqrt{2/\tau_\ell} \xi(t)$, such that $\langle \eta_\ell(t)\eta_\ell(t') \rangle = 2\sigma_\ell^2/\tau_\ell\delta(t-t')$. We obtain for the steady-state auto-correlation using \eref{lyapunov} and \eref{autocor}:
\begin{align}
	\langle\delta \ell(\tau) \delta \ell(0) \rangle = \sigma_\ell^2 e^{-\tau/\tau_\ell}.
\end{align}

\subsection{Non-Markovian signal}
\label{sec:signals:nonmarkov}
Not all ligand concentration trajectories encountered by cells are expected to be Markovian. For example, \textit{E. coli} swims in its environment with a speed which exhibits persistence. This leads to an auto-correlation function for the concentrations' derivative which does not decay instantaneously \cite{mattingly_escherichia_2021}. To model such a persistent signal, we use the classical model of a particle in a harmonic well
\begin{equation}
	\elabel{ho}
	\begin{split}
		\delta \dot \ell &= v(t), \\
		\dot v &= -\omega_0^2 \delta \ell(t) - v(t)/\tau_v +\eta_v (t),
	\end{split}
\end{equation}
where $\omega_0  = \sqrt{k/m}$, with $k$ the spring constant and $m$ the mass of the particle, $\tv$ is a relaxation timescale, and $\eta_v (t) = \sigma_v \sqrt{2 / \tv}\xi(t)$, with $\xi(t)$, as used throughout, a Gaussian white noise process of unit variance, and $\sigma_v$ the standard deviation of $v$. If the signal would obey the fluctuation-dissipation relation, then $m\sigma_v^2 = k_{\text{B}}T$, but since the biochemical signal could very well be generated via an active process this relation may not hold.
This process can be expressed as a 2-dimensional OU-process with:
\begin{align}
	\mathcal{J} &= \begin{pmatrix} 0 && 1 \\ -\omega_0^2 && -1/\tv  \end{pmatrix},\\
	\mathcal{B} &=  \begin{pmatrix} 0 && 0 \\ 0 &&  \sigma_v\sqrt{2/\tau_v}  \end{pmatrix}.
\end{align}
We find for the covariance matrix, using \eref{lyapunov}:
\begin{equation}
	\elabel{covmatNM}
	\mathcal{C}= \begin{pmatrix} \sigma^2_\ell && \sigma_{\ell v} \\ \sigma_{\ell v} && \sigma^2_v \end{pmatrix} = \sigma^2_v\begin{pmatrix}  1/\omega_0^2 && 0 \\ 0 && 1 \end{pmatrix}.
\end{equation}
Using \eref{autocor} we obtain the auto-correlation matrix in the overdamped regime, $\tau_v^{-1}>2\omega_0$,
\begin{equation}
	\elabel{cormatNM}
	\begin{split}
		\mathcal{C}(\tau) &= \begin{pmatrix}  \langle\delta\ell(\tau) \delta\ell(0) \rangle && \langle\delta\ell(\tau) \delta v(0) \rangle \\[2ex]
			\langle \delta v(\tau) \delta\ell(0)\rangle &&  \langle \delta v(\tau) \delta v(0) \rangle \end{pmatrix}, \\[4ex]
		&= \begin{pmatrix}  \sigma^2_\ell e^{-\mu \tau/2}\left(\cosh(\rho \tau)+\frac{\mu}{2 \rho}\sinh(\rho \tau)\right) && \sigma_v^2 e^{-\mu \tau/2}\frac{1}{\rho}\sinh(\rho \tau)\\[2ex]
			-\sigma_v^2 e^{-\mu \tau/2}\frac{1}{\rho}\sinh(\rho \tau) && \sigma^2_v e^{-\mu \tau/2}\left(\cosh(\rho \tau)-\frac{\mu}{2\rho}\sinh(\rho \tau)\right)  \end{pmatrix},
	\end{split}
\end{equation}
where $\rho = \sqrt{\mu^2/4 - \omega_0^2}$, with $\mu=\tv^{-1}$. The range of ligand concentrations which \textit{E. coli} might encounter is very large, based on the dissociation constants of the inactive and active receptor conformations, which for the Tar-MeAsp receptor ligand combination respectively are $K_{\text{D}}^{\text{I}}=18\mathrm{\mu  M}$ and $K_{\text{D}}^{\text{A}}=2900 \mathrm{\mu M}$ \cite{sourjik_functional_2004, mello_effects_2007}. This suggests that the variance in the ligand concentration is very large relative to that of the derivative of the ligand concentration, which is set by the swimming behaviour of the cell. For this reason we specifically focus on the limit where $\omega_0\to 0$, which corresponds to a vanishingly small spring constant, or a harmonic potential which becomes extremely wide. The variance in the concentration $\sigma_\ell^2$ then diverges, the normalized correlation functions in this limit are

\begin{equation}
	\elabel{cormatNMlim}
	\lim_{\omega_0\to 0}\begin{pmatrix}  \frac{\langle\delta\ell(\tau) \delta\ell(0) \rangle}{\sigma_\ell^2} && \frac{\langle\delta\ell(\tau) \delta v(0) \rangle}{\sigma_\ell \sigma_v} \\[2ex]
		\frac{\langle \delta v(\tau) \delta\ell(0)\rangle}{\sigma_\ell \sigma_v} &&  \frac{\langle \delta v(\tau) \delta v(0) \rangle}{\sigma_v^2} \end{pmatrix}
	= \begin{pmatrix} 1 && 0\\[2ex]
		0 && e^{-\mu \tau}  \end{pmatrix}.
\end{equation}

\section{Information bottleneck framework and solutions}
\label{sec:IBM}
Anticipating future environmental conditions allows for timely
adaptation. However, storing information costs resources such as
proteins, energy and time, and not all information in the past ligand
concentrations will be relevant for predicting the signal's future
state. Assuming that resources are in limited supply, this means that
cells must be efficient in which, and how much information they
store. This is elegantly captured in the Information Bottleneck
Method (IBM), which describes the problem of maximizing the information on the future signal while minimizing the information on the past signal that is stored in the network output, from which the future input is predicted \cite{tishby_information_1999}. The objective
function for the prediction of a variable of interest $ z_\tau \equiv z(t+\tau)$ is:
\begin{equation}
	\elabel{ibm}
	\max_{P(X_0|\vec{L}_p)}: \quad\mathcal{L}= I(x_0; z_\tau) - \gamma I(x_0; \vec{L}_p).
\end{equation}
The value of the sensing system output at the current time $t$ is $x_0 \equiv x(t)$. The variable of interest $z_\tau$ at a future time $t+\tau$ is the future concentration
$\ell_\tau \equiv \ell(t+\tau)$ for the Markovian signal, and the future
concentration derivative $v_\tau \equiv v(t+\tau)$ for the non-Markovian signal. Since the system of interest needs to predict one signal characteristic (either the future signal value or its derivative), one output component is sufficient for encoding the required information, as we describe in more detail below. The vector
$\vec{L}_p = (\delta \ell(0),\,\delta \ell(-\Delta t),\,\dots,\,\delta
\ell(-(N-1)\Delta t))^T$ is the past trajectory of ligand
concentrations of length $N$, discretized with timestep $\Delta
t$. The mutual information between the current system output and the
future property of interest is the predictive information
$I_\text{pred} \equiv I(x_0; z_\tau)$, and the mutual information between
the current system output and the past ligand concentration trajectory
is the past information $I_\text{past} \equiv I(x_0; \vec{L}_p)$. The
Lagrange multiplier $\gamma$ sets the relative cost of storing past
information over obtaining predictive information. Given a value of
$\gamma$, \eref{ibm} is maximized by optimizing the mapping of the
past ligand concentration trajectory $\vec L_p$ onto the current
output $x_0$. Since, by the data processing inequality, we have
$I_\text{past} \geq I_\text{pred}$, for $\gamma=1$ the objective
function is maximized by $I_\text{past} = I_\text{pred} =0$. As
$\gamma$ is decreased both the past and predictive information
increase, and the parametric curve in the $I_\text{past}-I_\text{pred}$
plane that arises is the information bound. For $\gamma=0$ there is no
cost to storing past information. The predictive information is then
only limited by the amount of information contained in the past about
the future signal property: $I_\text{pred} \leq I(\vec L_p ; z_\tau)$.

\subsection{Gaussian information bottleneck}
\label{sec:IBM:gaussian}
In general equation \ref{eqn:ibm} can be difficult to solve, as all mappings from $\vec L_p$ to $X_0$ are allowed. However, the problem becomes analytically tractable when the joint probability distribution of $\vec L_p$ and $z_\tau$ is a multivariate Gaussian. Here, we follow the procedure of Chechik and coworkers to obtain this mapping \cite{chechik_information_2005}. In the Gaussian model, the optimal mapping from $\vec L_p$ to $x_0$ is a linear one \cite{chechik_information_2005}
\begin{equation}
	\elabel{map}
	x_0 = \vec{A} \vec L_p + \xi; \quad \xi \sim N(0, \sigma^2_\xi),
\end{equation}
where $\vec{A}$ is a row vector which determines how strongly each entry in $\vec{L}_p$ contributes to the scalar output $X_0$ at any point in time. The random variable $\xi$ is the noise added to the signal due to the stochastic nature of the mapping; it is a Gaussian random variable independent of $\vec{L}_p$ with $0$ mean and variance $\sigma^2_\xi$. Finding the optimal mapping from $\vec{L}_p$ to $x_0$ corresponds to finding the optimal combination of $\vec{A}$ and $\sigma^2_\xi$. It can be shown that for any pair $(\vec{A}, \sigma^2_\xi)$, there exists a pair $(\vec{A}', 1)$ which yields the same values for $I_\text{past}$ and $I_\text{pred}$ after maximization of \eref{ibm} \cite{chechik_information_2005}. Therefore, we can set $\sigma^2_\xi=1$ without altering the information curve.

To obtain the information bound, we rewrite \eref{ibm} using the definition of the mutual information between Gaussian random variables:
\begin{equation}
	\elabel{lagrange}
	\begin{split}
		\mathcal{L} &= \frac{1}{2}\log(\sigma_x^2/\sigma_{x|z}^2) - \gamma \frac{1}{2}\log(\sigma_x^2/\sigma_{x|L}^2),
	\end{split}
\end{equation}
with the total variance $\sigma_x^2$ in the output $x_0$, the output variance conditional on the future signal property $\sigma_{x|z}^2\equiv \sigma^2_{x|z_\tau}$, and the output variance conditional on the complete history of ligand concentrations $\sigma_{x|L}^2\equiv\sigma^2_{x|\vec{L}_p}$. The latter is just the variance caused by the intrinsic noise,  $\sigma_{x|L}^2=\sigma_\xi^2=1$. The total variance in $x_0$ can be expressed in terms of the mapping vector $\vec A$ and the variance in the past signal using \eref{map}, $\sigma_x^2=\vec{A}\mat{\Sigma}_L \vec{A}^T + 1$, where $\mat{\Sigma}_L\equiv \mat{\Sigma}_{\vec{L}_p}$ is the covariance matrix of the past ligand concentration trajectory $\vec{L}_p$. To express the output variance conditional on the future signal property $z_\tau$ we use the Schur complement formula, which in general form reads:
\begin{equation}
	\elabel{schur}
	\mat\Sigma_{x|y} = \mat\Sigma_x - \mat\Sigma_{xy} \mat\Sigma_y^{-1} \mat\Sigma_{yx},
\end{equation}
where $\mat\Sigma_{yx} = \mat\Sigma_{xy}^T$. Using this formula to rewrite $\sigma^2_{x|z}$, and then using the linear relation from \eref{map} again, we obtain $\sigma^2_{x|z} = \vec{A} \mat\Sigma_{L|z}\vec{A}^T + 1$.

Filling in the expressions for the variances in $\mathcal{L}$ (\eref{lagrange}) gives:
\begin{equation}
	\begin{split}
		\mathcal{L} &= \frac{1}{2}\left((1-\gamma)\log \left(\left|\vec{A}\mat{\Sigma}_L\vec{A}^T + 1\right|\right)-\log \left(\left|\vec{A} \mat\Sigma_{L|z}\vec{A}^T + 1\right|\right)\right).
	\end{split}
\end{equation}
For any symmetric matrix $\mat C$ we have $\frac{\delta}{\delta A}\log |\vec A \mat C \vec A^T| = \left(\vec A \mat C \vec A^T\right)^{-1}2\vec A \mat C$, such that we obtain for the derivative of $\mathcal{L}$ to $\vec A$:
\begin{equation}
	\elabel{dL}
	\begin{split}
		\frac{\delta \mathcal{L}}{\delta\vec{A}} = (1-\gamma)\frac{\vec{A} \mat{\Sigma}_L}{\vec{A}\mat{\Sigma}_L\vec{A}^T + 1}-\frac{\vec{A} \mat{\Sigma}_{L|z}}{\vec{A}\mat{\Sigma}_{L|z}\vec{A}^T + 1}.
	\end{split}
\end{equation}
In our case $\vec{A}$ is a row vector, and both denominators are thus scalars. We find the maximum of $\mathcal{L}$ by equating its derivative to $0$, which gives:
\begin{equation}
	\elabel{eigprob}
	\begin{split}
		\vec{A} \mat{\Sigma}_{L|z} \mat{\Sigma}_L^{-1} &= (1-\gamma)\frac{\vec{A}\mat{\Sigma}_{L|z}\vec{A}^T + 1}{\vec{A}\mat{\Sigma}_L\vec{A}^T + 1}\vec{A}.
	\end{split}
\end{equation}
For this equality to hold $\vec{A}$ must either be identically 0, or a left eigenvector of the matrix $\mat{\Sigma}_{L|z} \mat{\Sigma}_L^{-1}$ with eigenvalue:
\begin{equation}
	\elabel{eigval}
	\lambda = (1-\gamma)\frac{\vec{A}\mat{\Sigma}_{L|z}\vec{A}^T + 1}{\vec{A}\mat{\Sigma}_L\vec{A}^T + 1}.
\end{equation} 
Here, we note that if the signal statistics is sufficiently rich and the prediction complexity sufficiently large (because, for example, multiple signal characteristics need to be predicted), then the matrix $\mat{\Sigma}_{L|z} \mat{\Sigma}_L^{-1}$ has multiple eigenvectors with non-trivial eigenvalues $0<\lambda_i<1$ \cite{chechik_information_2005}. This reflects the idea that storing the past information that is necessary to enable this complex prediction task may require multiple output components, i.e. an output vector $\vec{x}$, where each output component has an integration kernel given by one of the eigenvectors of $\mat{\Sigma}_{L|z} \mat{\Sigma}_L^{-1}$ \cite{chechik_information_2005}. However, for Markovian signals only one eigenvector with non-trivial eigenvalue $0<\lambda<1$ emerges, which means that one output component is sufficient to encode the required information. For the non-Markovian signals studied here,  $\mat{\Sigma}_{L|z} \mat{\Sigma}_L^{-1}$  has two eigenvectors if both the future value and its derivative need to be predicted (and $z=(\ell_\tau, v_\tau)$); to optimally predict both features from the current output, two output components are then required, provided $I_{\text{past}}$ is sufficiently large. However, here we consider the scenario that only the future derivative needs to be predicted, in which case only one non-trivial eigenvector emerges, and one output component is sufficient for encoding the required information. We leave the problem of predicting multiple signal features via multiple output components for future work.

We can define the optimal mapping $\vec A=||A||\vec\nu$ where $\vec \nu$ is the normalized left eigenvector of $\mat{\Sigma}_{L|z} \mat{\Sigma}_L^{-1}$ corresponding to its smallest eigenvalue,  $0<\lambda < 1$. The magnitude can be found by solving \eref{eigval} for $||A||$, using from \eref{eigprob} that $\lambda \vec \nu \mat{\Sigma}_L \vec \nu^T = \vec \nu \mat{\Sigma}_{L|z} \vec \nu^T$. This gives for the optimal mapping:
\begin{equation}
	\vec{A}^\text{opt} = \begin{cases} \sqrt{\frac{1-\gamma-\lambda}{\vec{\nu}_1 \mat{\Sigma}_L\vec{\nu}_1^T\lambda\gamma}}\vec{\nu}_1 \quad& \text{for} \quad 0<\lambda<1-\gamma, \\
		\hspace{30pt} 0 \quad& \text{for} \quad 1-\gamma \leq \lambda \leq 1.
	\end{cases}
\end{equation}

We can substitute $||A||^2 = (1-\gamma-\lambda)/(\vec{\nu} \mat{\Sigma}_L\vec{\nu}^T\lambda\gamma)$ in the definitions for the mutual information to express them in terms of $\lambda$ and $\gamma$. For the past information we obtain:
\begin{equation}
	\elabel{ibmpast}
	\begin{split}
		I_\text{past} &= \frac{1}{2}\log\left(||A||^2\vec \nu \mat{\Sigma}_L\vec \nu^T + 1\right),\\
		&= \frac{1}{2}\log\left(\frac{1-\gamma}{\gamma}\frac{1-\lambda}{\lambda}\right). \\
	\end{split}
\end{equation}
And for the predictive information:
\begin{equation}
	\elabel{ibmpred}
	\begin{split}
		I_\text{pred} &= \frac{1}{2}\log\left(||A||^2\vec \nu \mat{\Sigma}_L\vec \nu^T + 1\right) - \frac{1}{2}\log\left(||A||^2\vec \nu \mat{\Sigma}_{L|\ell_\tau}\vec \nu^T + 1\right),\\
		&= I_\text{past} - \frac{1}{2}\log\left(\frac{1-\lambda}{\gamma}\right), \\
		&= \frac{1}{2}\log\left(\frac{1-\gamma}{\lambda}\right).
	\end{split}
\end{equation}

\subsection{Markovian signal}
\label{sec:IBM:markov}
To obtain the information bound for prediction of the future ligand concentration of a Markovian signal, we need to determine the eigenvalues and vectors of the matrix (see \erefstwo{eigprob}{eigval})
\begin{equation}
	\mat W = \mat \Sigma_{L | \ell_\tau}\mat{\Sigma}_L^{-1}.
\end{equation}
Using the Schur complement formula (\eref{schur}) to rewrite the conditional matrix gives $\mat \Sigma_{L | \ell_\tau} = \mat{\Sigma}_L-  \mat{\Sigma}_{L \ell_\tau} \mat{\Sigma}_{L \ell_\tau}^T/\sigma_\ell^2$. Then defining the normalized matrices $\mat{R}_L = \mat{\Sigma}_L/\sigma_\ell^2$ and $\mat{R}_{L \ell_\tau}=\mat{\Sigma}_{L \ell_\tau}/\sigma_\ell^2 $ we find
\begin{equation}
	\begin{split}
		\elabel{Wmat}
		\mat W = \mathbb{I}_N-\mat{R}_{L \ell_\tau}\mat{R}_{L \ell_\tau}^T\mat{R}_L^{-1}.
	\end{split}
\end{equation}
where $N$ is the length of the input trajectory $\vec L_p$. The correlation matrix of the past trajectory is symmetric with entries $\mat{R}_L^{(i,j)}=\exp(-|i-j|\Delta t/\tau_\ell)$, where $\Delta t$ is the discretization timestep of the past trajectory $\vec L_p$ and $i$ ranges from $1$ to $N$. This is a Kac-Murdock-Szegö matrix, and its inverse is known:
\begin{equation}
	\mat{R}_L^{-1} =\frac{1}{1-e^{2\Delta t/\tau_\ell}} 
	\begin{pmatrix} 
		1       &   -e^{-\Delta t /\tau_\ell}   &              0                &  \dots    & \dots     &   0 \\
		-e^{-\Delta t /\tau_\ell}   &   1 + e^{-2\Delta t /\tau_\ell}    &   -e^{-\Delta t /\tau_\ell}   &   \dots   &   \dots   &   0 \\
		0       &   -e^{-\Delta t /\tau_\ell}   & 1 + e^{-2\Delta t /\tau_\ell}  & \ddots    & \dots     &   0 \\
		\vdots  &   \vdots                      & \ddots                        & \ddots    & \ddots    & \vdots \\
		0       &   \dots   & \dots     & - e^{-\Delta t /\tau_\ell}    & 1 + e^{-2\Delta t /\tau_\ell}  & -e^{-\Delta t /\tau_\ell} \\
		0       &   \dots   & \dots     & 0                             & -e^{-\Delta t /\tau_\ell}     & 1 \end{pmatrix}.
\end{equation}
Note that the inverse matrix is tridiagonal. The length $N$ cross-correlation vector between past trajectory and future concentration has entries $\mat{R}_{L \ell_\tau}^{(i)}=\exp(-(\tau+(i-1)\Delta t)/\tau_\ell)$. The product of the correlation matrices is surprisingly simple:
\begin{equation}
	\mat{R}_{L \ell_\tau}\mat{R}_{L \ell_\tau}^T\mat{R}_L^{-1} = e^{-2\tau/\tau_\ell} \begin{pmatrix}
		1                               &   0    &  \dots   &   0 \\
		e^{-\Delta t /\tau_\ell}        &   0    &  \dots   &   0 \\
		\vdots                          & \vdots &  \ddots  &   \vdots \\
		e^{-(N-1)\Delta t /\tau_\ell}  &   0    &  \dots   &   0
	\end{pmatrix}.
\end{equation}
Using this result we can straightforwardly determine the eigenvalues,
\begin{equation}
	\begin{split}
		&\left|\mat W - \lambda \mathbb{I}_N\right| = 0, \\
		\\
		&\left|\begin{pmatrix}
			1-\lambda-e^{-2\tau/\tau_\ell}      &   0           &  \dots   &   0 \\
			-e^{-(\tau+\Delta t)/\tau_\ell}     &   1-\lambda   &  \dots   &   0 \\
			\vdots                              &   \vdots      &  \ddots  &  \vdots \\
			-e^{-(\tau+(N-1)\Delta t)/\tau_\ell}&   0           &  \dots   &   1-\lambda
		\end{pmatrix}\right|=0.
	\end{split}
\end{equation}
The only contribution to the determinant comes from the diagonal, and the only nontrivial eigenvalue is thus $\lambda=1-e^{-2\tau/\tau_l}$. The optimal mapping is thus onto a one-dimensional scalar output $x_0$. The corresponding left eigenvector is given by
\begin{equation}
	\vec \nu_1 \mat W = (1-e^{-2\tau/\tau_l})  \vec \nu_1,
\end{equation}
which holds for $\vec \nu_1 = \begin{pmatrix}1 & 0 & \dots & 0 \end{pmatrix}$. The optimal mapping for the prediction of a one-dimensional OU-process is thus to copy its most recent value. This agrees with intuition as for any Markovian process, all the information about the future signal is contained in the most recent value. For a continuous input signal (rather than a discretized signal), and a continuous integration kernel $k(t)$ (rather than a mapping vector $\vec{A}$), this means that the optimal integration kernel is $k^{\text{opt}}(t) = a \delta (t)$.

\subsection{Non-Markovian signal}
\label{sec:IBM:nonmarkov}
To find the optimal mapping for the prediction of the derivative of a non-Markovian signal, based on its history of ligand concentrations, we need to find the eigenvalues and vectors of the matrix
\begin{equation}
	\begin{split}
		\mat W &= \mat \Sigma_{L | v_\tau}\mat{\Sigma}_L^{-1},\\
		&= \mathbb{I}_N - \frac{1}{\sigma_v^2}\mat \Sigma_{L v_\tau}\mat \Sigma_{L v_\tau}^T\mat \Sigma_{L}^{-1}.
	\end{split}
\end{equation}
The covariance matrix of the past trajectory is symmetric with entries $\mat \Sigma_{L}^{(i,j)}=\langle \delta \ell(0)\delta \ell(|i-j|\Delta t)\rangle$ where both $i$ and $j$ range from $1$ to $N$, the past trajectory length. The covariance vector between past trajectory and future derivative has entries $\mat{\Sigma}_{L v_\tau}^{(i,j)}= \langle \delta \ell(0)\delta v(\tau+(i-1)\Delta t)\rangle$. Both the concentration auto-correlation function, and the concentration to future derivative cross-correlation function, are shown in \eref{cormatNM}.

To better understand the optimal mapping of this signal we numerically investigate the eigenvalues of the matrix $\mat W$. For the prediction of $v_\tau$, there is only one non-trivial eigenvalue. Like for the Markovian signal, this shows that for the prediction of the derivative of this non-Markovian signal, the optimal mapping is always onto a scalar output. The non-trivial eigenvalue $\lambda$ decreases with the discretization timestep $\Delta t$ and is minimal for $\Delta t\to0$ \fref{Ndt}. In this limit, $\lambda$ has the same magnitude for any $N\geq2$, see \fref{Ndt}. A smaller eigenvalue $\lambda$ corresponds to larger past and predictive information and a larger ratio $I_\text{pred}/I_\text{past}$ (\eref{ibmpast} and \eref{ibmpred}), given any value of the Lagrange multiplier $\gamma$. For the optimal mapping we must thus have $N\geq2$ and $\Delta t \to 0$, where $N$ sets both the past trajectory and the mapping vector length. Because increasing the length above two does not yield an improvement in the value of $\lambda_1$ we focus on $N=2$.

\begin{figure}[tbhp]
	\centering
	\includegraphics[width=0.35\linewidth]{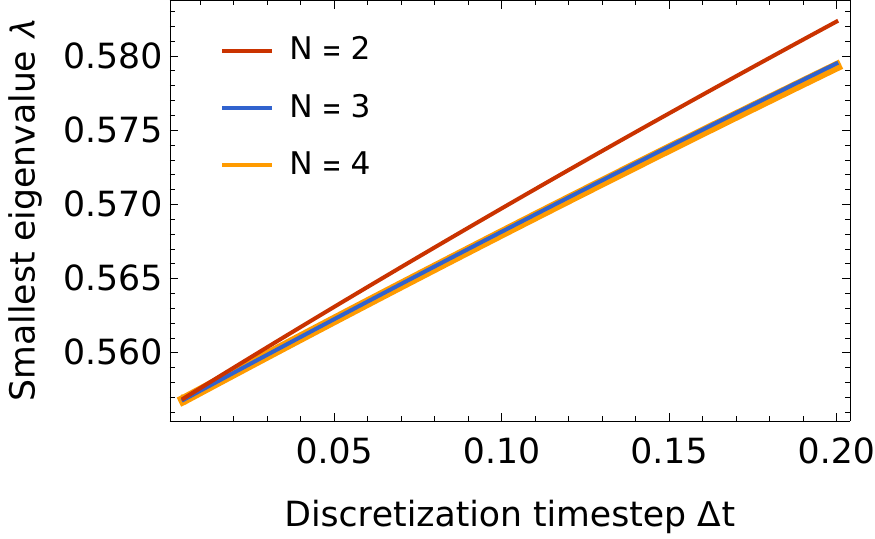}
	\caption{\textbf{The smallest eigenvalue of the IB matrix is minimal for $N\geq 2$ and $\Delta t \to 0$.} A smaller eigenvalue corresponds to a larger ratio $I_\text{pred}/I_\text{past}$ for any given value of the Lagrange multiplier $\gamma$. Parameters: friction timescale $\tau_v^{-1} =0.862 s^{-1}$ as determined in \cite{mattingly_escherichia_2021}, prediction interval $\tau=\tau_v$, and $\omega_0=0.4 s^{-1}$ such that the system is slightly overdamped.}
	\label{fig:Ndt}
\end{figure}

The fact that to reach the optimum we must have $N=2$ and $\Delta t \to 0$, shows that the optimal kernel $A$ takes an instantaneous measurement of a combination of the most recent ligand concentration, and its derivative. This can be understood as follows, for a trajectory of length two, the mapping vector also has length two, $\vec A = ||A||(\hat w_1, \hat w_2)$, with $\sqrt{\hat w_1^2+ \hat w_2^2}=1$. We can then express the linear mapping of $\vec L_p$ to $x_0$ (\eref{map}) as:
\begin{equation}
	\elabel{concdermap}
	x_0 =||A||\left[ (\hat w_1+\hat w_2)\delta \ell(0) - \hat w_2 \Delta t \frac{\delta \ell(0)-\delta \ell(-\Delta t)}{\Delta t}\right] + \xi,
\end{equation}
This expression shows that, as $\Delta t \to 0$, the two entries of $\vec A$ combine both the most recent signal value and the most recent derivative to generate $x_0$. This is intuitive because the signal is completely defined by its concentration and derivative (\eref{ho}). For this reason, and to obtain analytical insight into the optimal weights, we inspect the final two entries of the past ligand concentration trajectory in the limit $\Delta t \to 0$, which defines the past signal in terms of its most recent concentration and derivative
\begin{equation}
	\vec S_0 \equiv \begin{pmatrix}
		\delta \ell(0) & v(0) 
	\end{pmatrix}^T.
\end{equation}
Because the signal is Markovian in the joint properties $\delta \ell$ and $v$, the vector $\vec S_0$ contains the same information as the trajectory $\vec L_p$. The past information is now the mutual information between $x_0$ and $\vec S_0$, i.e. $I_\text{past}=I(x_0;\vec S_0)$. The output $x_0$ can then also be written as a projection of $\vec S_0$ via the alternative mapping vector $\vec{\tilde{A}} = ||A||(\hat a, \hat b)$:
\begin{equation}
	\elabel{concdermap2}
	\begin{split}
		x_0 =||A||\left(\hat a \delta \ell(0) +\hat b v(0)\right) + \xi.
	\end{split}
\end{equation}
Comparison with \eref{concdermap} shows how the components of $\vec{\tilde{A}}$ relate back to those in $\vec A$,
\begin{align}
	\hat w_1 &= \hat a +\hat b/ \Delta t, \\
	\hat w_2&= -\hat b/ \Delta t.
\end{align}

To obtain the optimal mapping vector $\vec{\tilde{A}}$ the matrix of signal statistics of which the eigenvalues and -vectors need to be determined is
\begin{equation}
	\mat W = \mat\Sigma_{\vec s|v_\tau}\mat{\Sigma_s}^{-1},
\end{equation}
with
\begin{align}
	\mat{\Sigma_s}&= \begin{pmatrix} \sigma^2_\ell && 0 \\ 0 && \sigma^2_v \end{pmatrix}, \\
	\mat\Sigma_{\vec s|v_\tau} &= \mat{\Sigma_s} - \frac{1}{\sigma_v^2}\mat\Sigma_{\vec s v_\tau}\mat\Sigma_{\vec s v_\tau}^T ,\\
	\mat\Sigma_{\vec s v_\tau} &= \begin{pmatrix}
		\langle \delta \ell(0) \delta v(\tau) \rangle \\
		\langle \delta v(0) \delta v(\tau) \rangle
	\end{pmatrix}.
\end{align}
We thus obtain
\begin{equation}
	\mat W = \mathbb I - \begin{pmatrix}
		\frac{\langle \delta \ell(0) \delta v(\tau) \rangle^2}{\sigma_\ell^2 \sigma_v^2} &  \frac{\langle \delta \ell(0) \delta v(\tau) \rangle \langle \delta v(0) \delta v(\tau) \rangle}{\sigma_v^4}\\
		\\
		\frac{\langle \delta \ell(0) \delta v(\tau) \rangle \langle \delta v(0) \delta v(\tau) \rangle}{\sigma_\ell^2\sigma_v^2} & \frac{\langle \delta v(0) \delta v(\tau) \rangle^2}{\sigma_v^4}
	\end{pmatrix}.
\end{equation}
This matrix has one nontrivial eigenvalue, $\lambda = 1-\frac{\langle \delta v(0) \delta v(\tau) \rangle^2}{\sigma_v^4}-\frac{\langle \delta \ell(0) \delta v(\tau) \rangle^2}{\sigma_\ell^2 \sigma_v^2}$, which depends on the normalized correlation functions between on the one hand the current concentration or derivative, and on the other hand the future derivative. The corresponding left eigenvector is
\begin{equation}
	\elabel{optvec}
	\vec \nu_1 = Q^{-1}\begin{pmatrix}
		\frac{1}{\sigma_\ell}\frac{\langle \delta \ell(0) \delta v(\tau) \rangle}{\sigma_\ell\sigma_v} & \frac{1}{\sigma_v}\frac{\langle \delta v(0) \delta v(\tau) \rangle}{\sigma_v^2}
	\end{pmatrix},
\end{equation}
where $Q$ normalizes the vector.Using the linear mapping $x_0 = ||A||\vec\nu_1\vec S_0 +\xi$, and defining $G\equiv ||A||/Q$, shows that the optimal output should be generated as follows
\begin{equation}
	\begin{split}
		x_0^\text{opt} &=G\left(\frac{\langle \delta \ell(0) \delta v(\tau) \rangle}{\sigma_\ell\sigma_v}\frac{\delta \ell(0)}{\sigma_\ell} + \frac{\langle \delta v(0) \delta v(\tau) \rangle}{\sigma_v^2}\frac{v(0)}{\sigma_v}\right) + \xi.
	\end{split}
\end{equation}
Clearly, the optimal mapping depends on the (normalized) cross-correlation coefficient $\rho_{\ell_0v_\tau} \equiv \avg{\delta \ell(0)\delta v(\tau)}/(\sigma_\ell \sigma_v)$  between the current concentration $\delta \ell(0)$ and future derivative $\delta v(\tau)$, and the cross-correlation coefficient $\rho_{v_0v_\tau}$ between the current derivative $\delta v(0)$ and future derivative $\delta v(\tau)$. Indeed, to optimally predict the future derivative, the cell should also use the current concentration and not only its current derivative. However, in the limit that the range of concentrations sensed becomes very large, corresponding to $\omega_0\to 0$, the current concentration is no longer correlated with the future derivative, and $\rho_{\ell_0v_\tau}\to 0$ (\eref{cormatNMlim}). In this limit, $\hat a =0$ and $\hat b=1$, and the kernel becomes a perfectly adaptive, derivative-taking kernel:
\begin{equation}
	\begin{split}
		\lim_{\omega_0\to 0} x_0^\text{opt} &=||A|| v(0) + \xi.
	\end{split}
\end{equation}
If we translate this back to the vector $ ||A||(\hat w_1, \hat w_2)$, operating on a ligand concentration trajectory $\vec L_p$, the optimal weights become $\hat w_1=-\hat w_2$.


\section{Past and predictive information for linear signalling networks}
\label{sec:IpastIpred}
In order to address how close biochemical networks can come to the information bounds derived above, we here describe how we obtain the past and predictive information for any linear (biochemical) network. We then use the resulting general expressions to compute the past and predictive information for the push-pull network and the chemotaxis system of the main text.

For any  linear network the output can be written as
\begin{equation}
	\delta x(t) = \int_{-\infty}^t ds\, k(t-s) \delta\ell(s) + \eta_x(t).
	\elabel{dx}
\end{equation}
The mapping kernel $k(t)$ is a property of the network and describes how the input signal is mapped onto the output. The noise term $\eta_x(t)$ is a sum of convolutions over all white noise processes in the network and corresponding network mapping functions, see \eref{timesol}. The variance in the output can generally be split up in a part caused by the signal and a part cause by the noise, and we have
\begin{equation}
	\begin{split}
		\sigma_x^2 &= \int_{-\infty}^t ds\int_{-\infty}^t ds' k(t-s) k(t-s') \langle \delta \ell(s)\delta \ell(s')\rangle  + \sigma_{\eta_x}^2, \\
		&= \sigma_{x|\eta}^2 + \sigma_{x|L}^2,
	\end{split}
\end{equation}
where $\sigma_{x|\eta}^2$ is the signal variance, i.e. all noise terms are fixed, and $ \sigma_{x|L}^2$ is the noise variance, i.e. the complete history of the signal is fixed. Using this decomposition we find for the past information, which is the mutual information between the current output and the complete signal history,
\begin{equation}
	\elabel{ipastdef1}
	\begin{split}
		I_\text{past}(x_0; \vec L_p) &= \frac{1}{2} \log\left(\frac{\sigma_x^2}{\sigma_{x|L}^2}\right) = \frac{1}{2} \log(1+\text{SNR}),
	\end{split}
\end{equation}
where the signal-to-noise ratio is defined as  $\text{SNR}=\sigma_{x|\eta}^2/\sigma_{x|L}^2$. Using the same definition for the mutual information when deriving the predictive information between current output and future ligand concentration, we obtain
\begin{equation}
	\elabel{ipreddef1}
	\begin{split}
		I_\text{pred}(x_0; \ell_\tau) &= \frac{1}{2} \log\left(\frac{\sigma_x^2}{\sigma_{x|\ell_\tau}^2}\right),\\
		&= \frac{1}{2} \log\left(1+\frac{\sigma_{x|\eta}^2}{\sigma_{x|L}^2}\right)-\frac{1}{2} \log\left(1+\frac{\sigma_{x|\eta}^2- \langle\delta x(0) \delta \ell(\tau)\rangle^2/\sigma_\ell^2}{\sigma_{x|L}^2}\right), \\
		&= I_\text{past} - \frac{1}{2} \log(1+\text{cSNR}).
	\end{split}
\end{equation}
In the second line we used the Schur complement formula, \eref{schur},  to decompose the variance in the output conditioned on the future signal: $\sigma_{x|\ell_\tau}^2 = \sigma_{x}^2 - \langle\delta x(0) \delta \ell(\tau)\rangle^2/\sigma_\ell^2$. The quantity $\sigma_{x|\eta}^2- \langle\delta x(0) \delta \ell(\tau)\rangle^2/\sigma_\ell^2$ can be understood as follows: the first term $\sigma_{x|\eta}^2$ is the contribution to the total variance of the output $\sigma^2_x$ that comes from the signal variations, while the second term quantifies the variance in the output that is correlated with the future input. The difference is thus the variance in the output coming from the signal variations that are not correlated with the future input. The ratio in the second logarithm can thus be understood as a conditional SNR that quantifies the part of the signal to noise ratio that does \emph{not} contain information about the future signal. This becomes more clear when considering its form in terms of the mapping kernel and signal correlation functions. For any linear signalling network we have
\begin{equation}
	\sigma_{x|\eta}^2-\langle\delta x(0) \delta \ell(\tau)\rangle^2/\sigma_\ell^2 = \int_{-\infty}^0 ds\int_{-\infty}^0 ds' k(-s) k(-s')\left( \langle \delta\ell(s)\delta\ell(s')\rangle-\frac{\langle \delta\ell(\tau)\delta\ell(s')\rangle\langle \delta\ell(\tau)\delta\ell(s)\rangle}{\sigma_\ell^2}\right),
\end{equation}
where the term in parentheses is the conditional variance in the past signal trajectory given a future value, $\mat \Sigma_{L |\ell_\tau}$. The form in \eref{ipreddef1} thus tells us that the predictive information is equal to the past information, minus the bits that do not contain information about the future ligand concentration. This difference is indeed the part of the past information that does contain predictive information about the future signal.

Although the expression above (\eref{ipreddef1}) nicely relates the past and predictive information, a more straightforward way of obtaining the predictive information is by expressing it directly in terms of the correlation between the current output and the future ligand concentration:
\begin{equation}
	\elabel{iprednew}
	\begin{split}
		I_\text{pred}(x_0;\ell_\tau) &= \frac{1}{2} \log\left(\frac{\sigma_x^2}{\sigma_{x|\ell_\tau}^2}\right)= -\frac{1}{2} \log\left(1-\frac{\langle\delta x(0) \delta \ell(\tau)\rangle^2}{\sigma_{x}^2\sigma_{\ell}^2}\right),
	\end{split}
\end{equation}
where we again used the Schur complement formula to rewrite $\sigma_{x|\ell_\tau}^2$.  Written this way we thus see that the predictive information depends on the normalized correlation between the current network output and the future ligand concentration. We can simply exchange the future ligand concentration for the future derivative when considering the chemotaxis network. 

To compute the past information for linear signalling networks we use \eref{ipastdef1}, and we thus need to compute the SNR. To compute the predictive information for the prediction of a future ligand concentration, we need to compute the `future correlation function' $\langle\delta x(0) \delta \ell(\tau)\rangle$. For the prediction of the future derivative we need $\langle\delta x(0) \delta v(\tau)\rangle$.

\section{Push-pull network}
\label{sec:PPN}
We consider a push-pull network that consists of a
phosphorylation-dephosphorylation cycle downstream of a
receptor. When bound to ligand, the receptor itself or its
associated kinase, such as CheA in {\it E. coli}, catalyzes the
phosphorylation of a readout protein $x$, like CheY. Active readout
molecules $x^\ast$ can decay spontaneously or be deactivated by an
enzyme (phosphatase), such as CheZ in {\it E. coli}. This cycle is
driven by the turnover of fuel such as ATP. We recognize that inside
the living cell, the chemical driving is typically large: for
example, the free energy of ATP hydrolysis is about $20 k_{\text{B}}T$,
which means that the system essentially operates in the irreversible
regime \cite{govern_optimal_2014, malaguti_theory_2021}. This system
consists of the following reactions:
\begin{align}
	\ce{R + L &<=>[k_+][k_-] RL} \\
	\ce{RL + x &->[\kf] RL + x^\ast} \\
	\ce{X^* &->[\kr] X}
\end{align}
Both the total number of receptors $\RT=R+RL$ and read-out molecules $\xT=X+X^*$ are conserved moieties. The chemical Langevin equations of this system are:
\begin{align}
	\elabel{PPNRL}
	\dot{RL} &=  [R_T-RL(t)]\ell(t) k_+ -  RL(t) k_- +  B_c(RL, \ell) \xi_c(t), \\
	\elabel{PPNx}
	\dot x^* &= [X_T-x^*(t)]RL(t) \kf - x^*(t) \kr + B_x(RL, x^\ast) \xi_x(t),
\end{align}
where $RL$ is the number of bound receptors, $x^*$ the number of phosphorylated read-out molecules, and $\xi_i$ denote independent Gaussian white noise with unit variance, $\avg{\xi_i (t) \xi_j(t^\prime)} = \delta_{ij}\delta (t-t^\prime)$. The noise strengths are $B_c(RL, \ell) = \sqrt{(R_T-RL(t))\ell(t) k_+ +  RL(t) k_-}$ and $B_x(RL,x^*) = \sqrt{(X_T-x^*(t))RL(t) \kf + x^\ast(t) \kr}$. The steady-state fraction of ligand-bound receptors is $p\equiv \overline{RL}/R_T = \bar \ell /(\bar \ell + \KD)$ with the dissociation constant $\KD = k_-/k_+$, and the steady-state fraction of phosphorylated readout molecules is $f\equiv\bar{x}^\ast/\xT = p \RT /(p \RT + \kr/\kf)$.

In the linear-noise approximation, expanding \erefstwo{PPNRL}{PPNx} to first order around their steady state, the equations become
\begin{align}
	\delta\dot{RL} &= b \,\delta \ell(t)-\delta RL(t)/\tc + \eta_c(t), \\
	\delta\dot{x}^* &=\gamma \,\delta RL(t) - \delta x^*(t)/\tr + \eta_x(t). 
\end{align}
The parameters $b=\RT p(1-p)/(\bar\ell \tc)$ and $\gamma=\xT f(1-f)/(\RT p \tr)$ are effective rates of receptor-ligand binding and readout phosphorylation, respectively. The decay rate of correlations in the receptor-ligand binding state is $\tc^{-1}=\bar \ell k_+ + k_-$, and that of the readout phosphorylation state is $\tr^{-1}=p \RT \kf+\kr$. The rescaled white noise processes have strengths $\langle \eta_c^2 \rangle = B_c^2 =2 \RT p(1-p)/\tc$ and $\langle \eta_x^2\rangle = B_x^2 = 2 \xT f(1-f)/\tr$. 

\subsection{Model statistics}
The relevant quantity to compute the past information is the variance in the output, decomposed into the part caused by signal variation and the part caused by noise. To compute the predictive information we further need the correlation function between the current output and a future ligand concentration $\langle\delta \ell(\tau) \delta  x^*(0) \rangle$. These quantities can be obtained via their Fourier transforms, as in \eref{powspecgeneral} and \eref{crosspowspec}. The matrices describing the properties of the signalling network are, as defined below \eref{signallingOU},
\begin{align}
	\mathcal{G} &= \begin{pmatrix} b \\ 0 \end{pmatrix}, \\
	\mathcal J &= \begin{pmatrix}
		-\tc^{-1} & 0 \\ \gamma & -\tr^{-1}
	\end{pmatrix}, \\
	\mathcal{B}&=  \begin{pmatrix}
		\sqrt{\langle \eta_c^2 \rangle} & 0 \\ 0 &\sqrt{\langle \eta_x^2 \rangle}
	\end{pmatrix} =  \begin{pmatrix}
		\sqrt{2 \RT p(1-p)/\tc} & 0 \\ 0 & \sqrt{2 \xT f(1-f)/\tr}
	\end{pmatrix}.
\end{align}

A useful property of the network is the matrix exponential of its Jacobian, which in Fourier space is (see \eref{timesol} and \eref{fouriersol})
\begin{equation}
	\begin{split}
		\mathcal{F}\{e^{\mathcal J t}\}(\omega) &= (i\omega \mathbb{I}_2 - \mathcal{J})^{-1}, \\
		&= \begin{pmatrix}
			\frac{1}{1/\tc+i\omega} & 0\\ \\
			\frac{\gamma}{(1/\tc+i\omega)(1/\tr+i\omega)} & \frac{1}{1/\tr+i\omega}
		\end{pmatrix}. 
	\end{split}
\end{equation}
We then have $\mathbb{G}(\omega) =  \mathcal{F}\{e^{\mathcal J t}\}(\omega)\mathcal{G}$ and $\mathbb{N}(\omega) =  \mathcal{F}\{e^{\mathcal J t}\}(\omega)\mathcal{B}$, see also \eref{fouriersol} and \eref{powspecgeneral}. The integration kernel that maps the ligand concentration onto the output of the push-pull network, see \eref{dx}, is given by the inverse Fourier transform of the second entry of $\mathbb{G}(\omega)$, which is the frequency dependent gain, $\tilde g_{\ell \to x}(\omega)$, from $\ell$ to $x$:
\begin{equation}
	\begin{split}
		k(t) &\equiv \mathcal{F}^{-1}\{\tilde g_{\ell \to x}(\omega)\} = b\gamma\tc\tr \frac{1}{\tr-\tc}\left(e^{-t/\tr}-e^{-t/\tc}\right), \\
		&= \xT f(1-f) (1-p)/\bar \ell \frac{1}{\tr-\tc}\left(e^{-t/\tr}-e^{-t/\tc}\right),
	\end{split}
\end{equation}
The so-called static gain of the network is the integral of this kernel over all time, $\bar g_{\ell \to x} \equiv \int_0^{\infty}k(t)dt=\xT f(1-f) (1-p)/\bar \ell $. This parameter quantifies how much a step change in the input concentration changes  the steady-state level of the output: $\bar g_{\ell \to x} = \partial \bar{x^\ast}/\partial \bar\ell$. We will use this parameter in the statistical quantities that follow. The static gain is also given by $\bar{g}_{\ell \to x} = \bar{g}_{\ell \to RL} \bar{g}_{RL \to x}$, with  $\bar{g}_{\ell \to RL} = p(1-p) \RT / \bar \ell$ the static gain from $\bar\ell$ to $RL$ and $\bar{g}_{RL \to x}=f(1-f) \xT / (p \RT)$ the static gain from $RL$ to $x^\ast$.

We model the Markovian ligand concentration as a 1-dimensional OU process \eref{Markovsig}, which has the following power spectrum
\begin{equation}
	S_\ell(\omega) = \langle|\delta \ell (\omega)|^2\rangle = \frac{2 \sigma_\ell^2/\tau_\ell}{1/\tau_\ell^2+\omega^2}.
\end{equation}
This yields the following expression for the power spectra (see \eref{powspecgeneral}):
\begin{align}
	\mathbb{G}(-\omega)S_\ell(\omega)\mathbb{G}(\omega)^T &= b^2\begin{pmatrix}
		\frac{1}{1/\tc^{2}+\omega^2} &  \gamma\frac{1}{1/\tr-i\omega}\frac{1}{1/\tc^{2}+\omega^2}\\ \\
		\gamma \frac{1}{1/\tr+i\omega}\frac{1}{1/\tc^{2}+\omega^2} &  \gamma^2\frac{1}{1/\tr^2+\omega^2}\frac{1}{1/\tc^{2}+\omega^2}
	\end{pmatrix} \frac{2 \sigma_\ell^2/\tau_\ell}{1/\tau_\ell^2+\omega^2}\\[4ex]
	|\mathbb{N}(\omega)|^2 &= \langle \eta_c^2 \rangle\begin{pmatrix}
		\frac{1}{1/\tc^{2}+\omega^2} &  \gamma\frac{1}{1/\tr-i\omega}\frac{1}{1/\tc^{2}+\omega^2}\\ \\
		\gamma \frac{1}{1/\tr+i\omega}\frac{1}{1/\tc^{2}+\omega^2} &  \gamma^2\frac{1}{1/\tr^2+\omega^2}\frac{1}{1/\tc^{2}+\omega^2} + \frac{\langle \eta_x^2 \rangle}{\langle \eta_c^2 \rangle} \frac{1}{1/\tr^2+\omega^2}
	\end{pmatrix}
\end{align}
We thus have for the power spectrum of the read-out:
\begin{equation}
	\begin{split}
		S_x(\omega) &= \tilde g^2_{\ell \to x}(\omega) S_\ell(\omega) + N^2_x(\omega) \\
		&=\frac{ 2 b^2\gamma^2 \sigma_\ell^2/\tau_\ell}{(1/\tr^2+\omega^2)(1/\tc^{2}+\omega^2)(1/\tau_\ell^2+\omega^2)} +  \frac{\gamma^2\langle \eta_c^2 \rangle}{(1/\tr^2+\omega^2)(1/\tc^{2}+\omega^2)} +  \frac{\langle \eta_x^2 \rangle}{1/\tr^2+\omega^2},
	\end{split}
\end{equation}
The variance in the read-out $\sigma^2_x = 1/(2\pi) \int_{-\infty}^\infty S_x(\omega)$ is hence given by 
\begin{equation}
	\elabel{varx}
	\begin{split}
		\sigma_x^2 &= \sigma_{x|\eta}^2 + \sigma_{x|L}^2\\
		&= \bar g^2_{\ell \to x} \frac{1+\tr/\tau_\ell + \tr/\tc}{(1+\tc/\tau_\ell)(1+\tr/\tau_\ell)(1+\tr/\tc)} \sigma_\ell^2 +\bar g_{RL \to x}^2 \RT p(1-p)\frac{1}{1+\tr/\tc}+\xT f(1-f),\\
		&=  \underbrace{\bar g^2_{\ell \to x} \frac{1+\tr/\tau_\ell + \tr/\tc}{(1+\tc/\tau_\ell)(1+\tr/\tau_\ell)(1+\tr/\tc)}}_{\text{dynamical gain}}  \sigma_\ell^2  +\xT f(1-f)\left(1+\bar g_{\ell \to x}\frac{\bar \ell}{\RT p}\frac{1}{1+\tr/\tc}\right), \\
	\end{split}
\end{equation}
where $\bar g_{RL \to x} = \gamma \tau_r =\xT f(1-f)/(\RT p)$ is the
static gain from the receptor to the readout. The expression above gives
insight into the role of the different network components in shaping the noise in the readout. It can be seen
that the contribution from the signal variance $\sigma^2_\ell$ to $\sigma^2_x$ is determined by the static gain $\bar{g}_{\ell \to x}^2$, which is
proportional to $\xT$, and a factor that only depends on ratios of timescales. Their
product is the dynamical gain, which decreases
monotonically with $\tr$. The intrinsic noise in the phosphorylation
state of the read-outs leads to the noise term $\xT f(1-f)$, which
cannot be averaged out. The noise arising from ligand binding and
unbinding increases with the static gain, but can be mitigated by
increasing the number of receptors or the integration time $\tr$. The
latter strategy is what we call time-averaging.

The signal-to-noise ratio $\text{SNR}=\sigma_{x|\eta}^2/ \sigma_{x|L}^2$ can straightforwardly be obtained from \eref{varx}.  This is the quantity that sets the magnitude of the past information, see \eref{ipastdef1}. To determine the predictive information we need to compute the correlation function from the current output to the future ligand concentration $\langle \delta x(0) \delta \ell(\tau) \rangle$. This requires the cross-spectrum from output to ligand concentration, which is given by (\eref{crosspowspec})
\begin{equation}
	\begin{split}
		\tilde g_{\ell \to x}(-\omega) S_\ell(\omega) = \frac{b\gamma}{(1/\tc-i\omega)(1/\tr-i\omega)}\frac{2 \sigma_\ell^2/\tau_\ell}{1/\tau_\ell^2+\omega^2}.
	\end{split}
\end{equation}
From this power spectrum we obtain the required correlation function by taking the inverse Fourier transfrom:
\begin{equation}
	\begin{split}
		\elabel{futcorppn}
		\langle \delta x(0) \delta \ell(\tau) \rangle &= \mathcal{F}^{-1}\{\tilde g_{\ell \to x}(-\omega) S_\ell(\omega)\}, \\
		&= \frac{\bar g_{\ell \to x} \sigma_\ell^2}{(1+\tc/\tau_\ell)(1+\tr/\tau_\ell)}e^{-\tau/\tau_\ell}.
	\end{split}
\end{equation}
This correlation function thus decays exponentially with the prediction interval $\tau$ at a rate $\tau_\ell^{-1}$, just as the signal auto-correlation. The (squared) correlation coefficient, which sets $I_\text{pred}$, is given by $ \langle \delta x(0) \delta \ell(\tau) \rangle^2/( \sigma_\ell^2 \sigma_x^2) = \rho_{\ell x}^2 e^{-2 \tau/\tau_\ell}$, with the (squared) instantaneous correlation coefficient (for convenience given as its inverse)
\begin{equation}
	\elabel{rho2ppn}
	\begin{split}
		\rho^{-2}_{\ell x} =\frac{\bar{\ell}^2}{\sigma_\ell^2}&\left(1+\frac{\tr}{\tau_\ell}\right)^2\left(1+\frac{\tc}{\tau_\ell}\right)^2\bigg(\frac{1}{X_T f(1-f)(1-p)^2}+\frac{1}{R_T p(1-p)(1+\tr/\tc)} 
		+\frac{\sigma^2_\ell}{\bar{\ell}^2}\frac{1+\tr/\tau_\ell+\tr/\tc}{(1+\tc/\tau_\ell)(1+\tr/\tau_\ell)(1+\tr/\tc)}\bigg).
	\end{split}
\end{equation}
When the right-hand-side is minimized, the correlation is thus maximized. This expression shows that increasing $\xT$ and $\RT$ always increases the instantaneous correlation coefficient, and that the fraction of phosphorylated readout molecules in steady state that maximizes the correlation coefficient is $f=1/2$.

\subsection{Past and predictive information of the push-pull network} \label{sec:PPN:IpastIpred} Using the quantitites computed above, we can determine both the past and the predictive information.  For the past information we use \eref{ipastdef1}, whith the SNR from \eref{varx}:
\begin{equation*}
	\text{SNR} =\sigma_{x|\eta}^2/ \sigma_{x|L}^2 =(1-p)\frac{\sigma_\ell^2}{\bar \ell^2} \frac{1+\tr/\tau_\ell + \tr/\tc}{(1+\tc/\tau_\ell)(1+\tr/\tau_\ell)(1+\tr/\tc)}\bigg/\left(\frac{1}{\xT f(1-f)(1-p)}+\frac{1}{\RT p(1+\tr/\tc)}\right).
\end{equation*}
The predictive information is a function of the correlation between the current output and the future ligand concentration, \eref{iprednew}. This correlation can be decomposed into the instantaneous correlation coefficient and  an exponential decay on the timescale of the ligand concentration fluctuations, \eref{futcorppn}. We thus obtain for the predictive information,
\begin{equation}
	\elabel{ipredppn}
	I_\text{pred}(x_0 ; \ell_\tau) = -\frac{1}{2}\log(1-\rho_{\ell x}^2 e^{-2\tau/\tau_\ell}) .
\end{equation}
The instantaneous correlation coefficient $\rho_{\ell x}^2$ is given in \eref{rho2ppn}. From \eref{ipredppn} it also becomes clear that while the value of the predictive information depends on the forecast interval $\tau$, the optimal design of the network that maximizes the predictive information, determined by the optimal ratio $\xT/\RT$, the optimal integration time $\tr$, and the optimal ligand-bound receptor fraction $p$, does not depend on the forecast interval $\tau$. 

\subsection{Optimal resource allocation}
\label{sec:PPN:optallocation}
Increasing the number of receptor or readout molecules always increases the precision with which the cell can predict a signal (see \eref{rho2ppn}). However, when the total resource pool is constrained, the cell has to choose whether it makes more receptors or more readout molecules. To find the optimal ratio of read-out to receptor molecules we, can use the $C=A \RT + B \xT$ to express $\xT$ and $\RT$ in terms of the total cost $C$ and the ratio $\xT/\RT$:
\begin{align}
	\xT = C\frac{\xT/\RT}{A+ B \xT/\RT}, \\
	\RT = C\frac{1}{A+ B \xT/\RT}.
\end{align}
The factors $A$ and $B$ set the cost of receptors and readout molecules, respectively. Substituting these expressions for $\xT$ and $\RT$ into the expression for the correlation coefficient  between the  output and ligand concentration (\eref{rho2ppn}), setting the derivative of the resulting expression with respect to $\xT/\RT$ to zero, and solving for $\xT/\RT$ gives
\begin{equation}
	\begin{split}
		\elabel{ppnalloc}
		(\xT/\RT)^\text{opt} &= \sqrt{\left(1+\frac{\tr}{\tc}\right)\frac{p}{1-p}\frac{1}{f(1-f)}\frac{A}{B}}, \\
		&= 2\sqrt{p/(1-p)}\sqrt{1+\tr/\tc},
	\end{split}
\end{equation}
where for the second line we used $A=B=1$ and $f=f^\text{opt}=1/2$. This is the optimal ratio of readout to receptor molecules in the push-pull network, given an integration time $\tr$ and a steady state fraction of ligand-bound receptors $p$. Perhaps surprisingly, this optimal ratio $(\xT/\RT)^{\text{opt}}$ maximizes, for a given $\tr$ and $p$,  not only the predictive information, but also the past information. This is because the ratio $\xT/\RT$ determines, together with $\tr$ and $p$,  the interval $\Delta$ for sampling the ligand-binding state of the receptor: when the ratio $\xT/\RT$ obeys \eref{ppnalloc}, the readout molecules sample each receptor molecule roughly once every correlation time: $\Delta \sim \tc$ \cite{govern_optimal_2014, malaguti_theory_2021}. \eref{ppnalloc} is thus a statement about optimally extracting the information that is encoded in the receptor-ligand binding history, both concerning the past information and the predictive information. This is illustrated in \fref{optallocation}.

\begin{figure}[tbhp]
	\centering
	\includegraphics[width=0.4\linewidth]{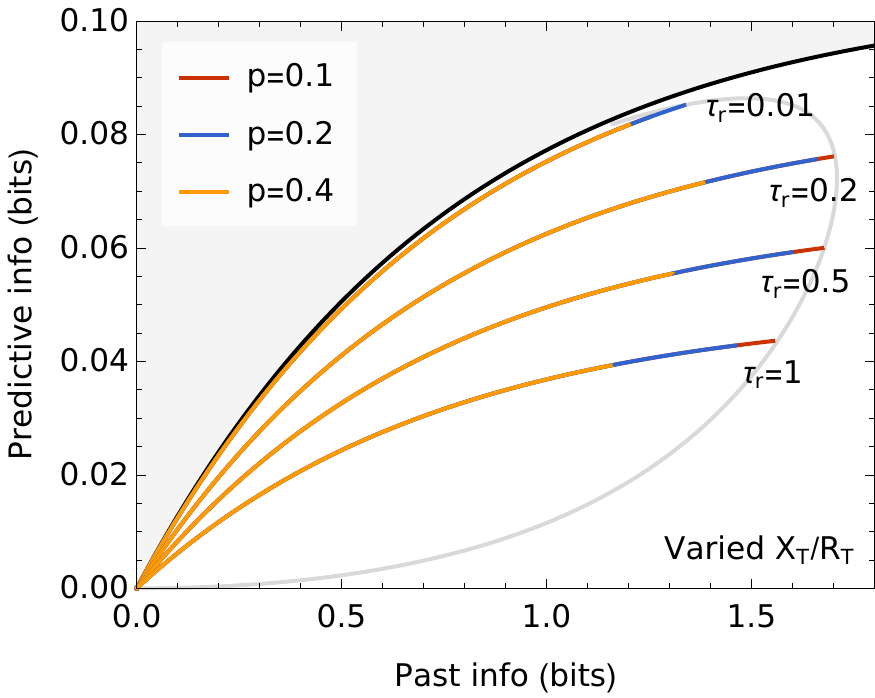}
	\caption{\textbf{The past and predictive information are maximized by the same ratio $\xT/\RT$ and fraction $p$.} The information plane, showing the information bound in black, and the isocost line $C=10^4$ in gray. To construct the coloured lines in this figure the ratio $\xT/\RT$ has been varied from zero to a value beyond the optimal value that maximizes $I_\text{past}$ and $I_\text{pred}$. This is done for several values of the receptor occupancy $p$ ($p=0.1$ in red, $p=0.2$ in blue, $p=0.4$ in orange), and for several values of $\tr$ (indicated in the figure). When $\xT/\RT$ reaches its optimal value, both $I_\text{past}$ and $I_\text{pred}$ are maximal. When the ratio is increased further the system moves back to the origin via the same coordinates. Only the integration time $\tr$ meaningfully distinuishes between strategies that maximize predictive or past information, or that approach the information bound. The reason is that $\xT/\RT$, together with $\tr$ and $p$, control the optimal extraction of information that is encoded in the receptor-ligand binding history, both concerning $I_{\text{past}}$ and $I_{\text{pred}}$. The gray isocost line is obtained by varying $\tr$, while maximizing for each $\tr$ the correlation coefficient given  by \eref{rho2ppn}; the latter is done by substituting \eref{ppnalloc} into \eref{rho2ppn} and numerically optimizing the resulting expression over $p$. The isocost line gives the region of $I_{\text{past}}$ and $I_{\text{pred}}$ that is accessible for a given resource cost $C$. Parameter values are $A=B=1$, $f=1/2$, $(\sigma_\ell/\bar\ell)^2=10^{-2}$, $\tc/\tL=10^{-2}$.}
	\label{fig:optallocation}
\end{figure}

\subsection{Operating costs diverge when approaching the information bound}
\label{sec:PPN:opcost}
The precision of any sensing device is limited by the resources that are devoted to it. The cost function we consider in this work is 
\begin{align}
	\elabel{SIcost}
	C=\lambda(\RT +\xT) + c_1 \xT \Delta \mu/\tr.
\end{align}
The first term is the maintenance cost; this is the cost of producing
new network components at the growth rate $\lambda$. The second term
is the operating cost and describes the chemical power that is
necessary to run the network; it depends on the flux through the
network, $\xT/\tr$, and the free-energy drop $\Delta \mu$ over a full
cycle of phosphorylation and dephosphorylation, given by the free energy of ATP
hydrolysis. The coefficient $c_1$ describes the relative energetic cost
of synthesising the components during the cell cycle, versus that of
running the system. In the main text we consider the case where
$c_1 \to 0$. Here we will investigate how close cells can come to the
information bound when $c_1$ is finite, thus including the chemical
power cost of running the network.

It is clear from $\eref{SIcost}$ that for finite $c_1$ the operating cost diverges when $\tr \to 0$. Because the optimal IBM solutions are instantaneous, this is precisely the limit in which the network must be to reach the information bound. As a consequence, when we consider the operating costs, the push-pull network can only be at the information bound when $(I_{\text{past}},I_{\text{pred}})\to (0,0)$ or $C \to \infty$ (\fref{opcost}A). The system can mitigate the operating costs by decreasing $\xT$, because this decreases the flux through the cycle. However, this also decreases the gain and thus, eventually, any information transduced through the network. In the limit that both $\xT$ and $\tr$ approach zero, the system approaches the information bound at the origin, see both \fref{opcost}A and B. More generally, when the running costs are taken into account, the system time averages more (i.e., $\tr$ rises), because frequent measurements are now even more costly. Still, $\tr$ decreases as the total resource availability $C$ grows.

\begin{figure}[tbhp]
	\centering
	\includegraphics[width=0.6\linewidth]{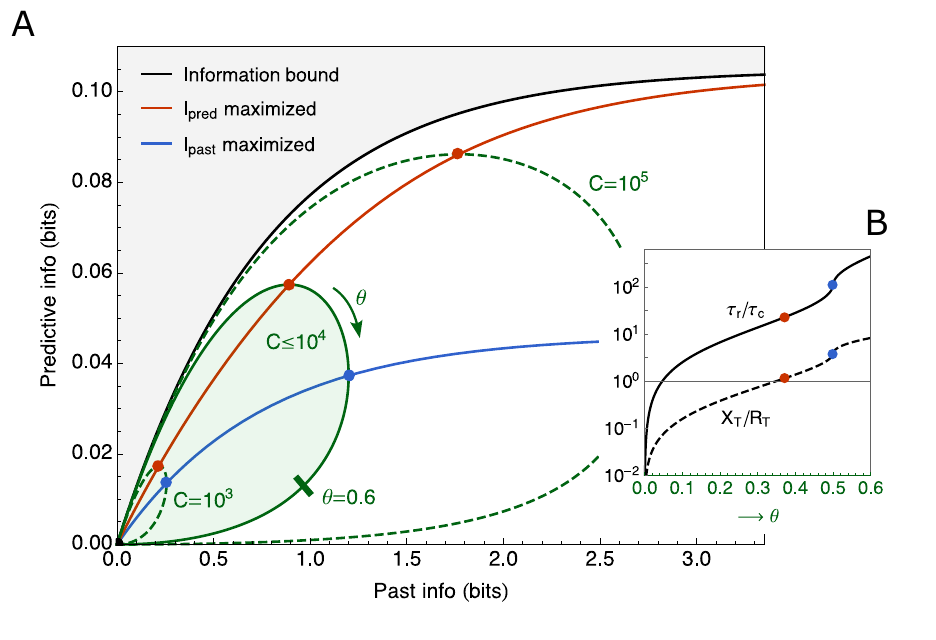}
	\caption{\textbf{Due to diverging operating costs the push-pull network only reaches the information bound for infinite resource availability.} (A)  In green, the region of accessible predictive and past information in the push-pull network under a resource constraint $C=\lambda(\RT +\xT) + c_1\xT \Delta \mu/\tr$, with $\lambda =1$ and $c_1 = 1/\Delta \mu$, corresponding to a cell doubling time of roughly $20 \text{min}$ \cite{govern_optimal_2014}. The black line is the information bound; the red and blue dots mark the points where $I_{\text{pred}}$ and $I_{\text{past}}$ are maximized, respectively, under a resource constraint $C$; the red and blue lines connect these points, respectively, for increasing $C$. The accesible region for $C\leq 10^4$ and the isocost lines for $C=10^3$ and $C=10^5$ have been obtained as described under \fref{optallocation}. The forecast interval has been set to one signal correlation time in the future: $\tau=\tau_\ell$. (B) The integration time over the receptor correlation time, $\tr/\tc$, and the ratio of the number of readout and receptor molecules, $\xT/\RT$, as a function of the distance $\theta$ along the iscocost line for $C=10^4$ in panel A. For $\theta \to 0$, both $\tr$ and $\xT$ go to zero, thus reducing both $I_{\text{past}}$ and $I_{\text{pred}}$ to zero. Other parameter values in both panels are $f=f^\text{opt}=1/2$, $(\sigma_{\ell} /\bar{\ell})^2 = 10^{-2}$, $\tc/\tL=10^{-2}$.}
	\label{fig:opcost}
\end{figure}

\section{Chemotaxis network}
\label{sec:CN}
The evidence is mounting that in the \textit{E. coli} chemotaxis system, receptors cooperatively control the activity of the kinase CheA \cite{maddock_polar_1993,Duke.1999,shimizu_modular_2010,Keegstra.2017}. Furthermore, the kinase activity is adaptive due to the methylation of inactive receptors \cite{segall_temporal_1986,parkinson_signaling_2015}. A widely used approach to describe the effects of receptor cooperativity and methylation on kinase activity, has been to employ the Monod-Wyman-Changeux (MWC) model \cite{monod_nature_1965,sourjik_functional_2004,mello_allosteric_2005,keymer_chemosensing_2006,mello_effects_2007,shimizu_modular_2010,kamino_adaptive_2020,mattingly_escherichia_2021}. We will follow this approach and, more specifically, model the chemotaxis system as described by Tu and colleagues \cite{tu_modeling_2008}. In this model, each receptor can switch between an active and inactive conformational state. Moreover, receptors are partitioned into clusters of equal size $N$. In the spirit of the MWC model, receptors within a cluster switch conformation in concert, so that each cluster is either active or inactive  \cite{monod_nature_1965}. Furthermore, it is assumed that receptor-ligand binding and conformational switching are faster than the other timescales in the system. The probability for the kinase, i.e. the receptor cluster, to be active, is then described by:
\begin{equation}
	a(\ell, m) = \frac{1}{1+\exp(\Delta F_T(\ell , m))},
\end{equation}
where $\Delta F_T(\ell, m)$ is the total free-energy difference between the active and inactive state, which is a function of the ligand concentration $\ell(t)$ and the methylation level of the cluster $m(t)$. The simplest model adopted here assumes a linear dependence of the total free-energy difference on the free-energy difference arising from ligand binding and methylation:
\begin{equation}
	\Delta F_T(\ell , m) = -\Delta E_0 + N(\Delta F_\ell(\ell) + \Delta F_m(m)),
\end{equation}
where the free-energy difference due to ligand binding is
\begin{equation}
	\Delta F_\ell(\ell) = \ln(1+\ell(t)/\KDI) -\ln(1+\ell(t)/\KDA).
\end{equation}
Between the two states the cluster has an altered dissociation constant, which is denoted $\KDI$ for the inactive state, and $\KDA$ for the active state. The free-energy difference due to methylation has been experimentally shown to depend approximately linearly on the methylation level \cite{shimizu_modular_2010}:
\begin{equation}
	\Delta F_m(m) = \tilde \alpha(\bar m -m(t)).
\end{equation}

We assume that inactive receptors are irreversibly methylated, and active receptors irreversibly demethylated, with zero-order ultrasensitive kinetics \cite{tu_modeling_2008,Emonet.2008,tostevin_mutual_2009}. The dynamics of the methylation level of the $i^\text{th}$ receptor cluster is then given by:
\begin{equation}
	\begin{split}
		\dot m_i =& (1-a_i(\ell, m_i)) k_R - a_i(\ell, m_i) k_B + B_{m_i}(a_i)\xi(t),
	\end{split}
\end{equation}
with $B_m^{(i)}(a_i) = \sqrt{(1-a_i(\ell, m_i)) k_R + a_i(\ell, m_i) k_B}$, and unit white noise $\xi(t)$. These dynamics indeed give rise to perfect adaptation, since from this equation we find that the steady state cluster activity is given by $p\equiv\bar a = 1/(1+k_B/k_R)$, thus indeed independent of the ligand concentration. 

Finally, active receptors catalyze phosphorylation of read-out molecules, and phosphorylated read-out molecules decay at a constant rate. We have
\begin{equation}
	\dot x^* = \sum^{\RT}_{i=1} a_i(t)(\xT-x^*(t)) \kf - x^*(t) \kr + B_x(a_i, x^\ast) \xi(t),
\end{equation}
where $\RT$ is the total number of receptor \emph{clusters}. The steady state fraction of phosphorylated read-outs is given by $f \equiv \bar{x}^\ast/\xT =(1+\kr/(\kf \RT p))^{-1}$.

\subsection{Linear dynamics}
We again do a first order approximation around the steady state, defining all variables in terms of deviations from their mean: $\delta \ell(t) = \ell(t)-\bar \ell$, $\delta m(t) = m(t)-\bar m$ and $\delta a(t) = a(t) - p$. The linear form of this model has previously been studied in for example \cite{tu_modeling_2008} and \cite{tostevin_mutual_2009}. We obtain for the linear dynamics of the $i^\text{th}$ cluster activity
\begin{equation}
	\elabel{dela}
	\delta a_i(t) = \alpha \delta m_i(t) - \beta \delta \ell(t),
\end{equation}
with $\alpha = \tilde \alpha N p(1-p)$ and $\beta =\kappa N p(1-p)$, with $\kappa= (\bar \ell +\KDI)^{-1}-(\bar \ell +\KDA)^{-1}$. For the methylation on the $i^\text{th}$ cluster and for the readout dynamics we then obtain, as a function of $\delta a(t)$,
\begin{align}
	\dot{\delta m_i} &= -\delta a_i(t)/(\alpha \tm) +\eta_{m_i}(t), \elabel{dmi}\\
	\dot{\delta x^\ast} &= \gamma \sum^{\RT}_{i=1} \delta a_i(t)-\delta x^*(t)/\tau_r+\eta_x(t),\elabel{dxcn}
\end{align}
where we have introduced the relaxation times $\tm=(\alpha(k_R+k_B))^{-1}$ for methylation and $\tr = (\RT p \kf+\kr)^{-1}$ for phosphorylation. We have further defined the rate at which an active cluster phosphorylates the readout CheY: $\gamma = \xT f(1-f)/(p \RT \tr)$. Substituting the expression for $\delta a_i$ in \eref{dela} into \erefstwo{dmi}{dxcn},  and expressing the dynamics in terms of the methylation on all clusters gives
\begin{align}
	\frac{d}{dt}\left(\sum^{R_T}_{i=1}\delta m_i \right) &= -\sum^{\RT}_{i=1}\delta m_i/\tm + q \delta \ell(t)/(\alpha \tm) + \eta_m(t), \\
	\dot{\delta x^\ast} &= -\delta x^\ast(t)/\tr - \gamma q  \delta \ell(t) + \gamma \alpha \sum^{\RT}_{i=1} \delta m_i(t)+ \eta_x(t),
\end{align}
with  $q =\RT \beta$ (see \eref{dela} for $\beta$). The rescaled white noise $\eta_m$ is the sum of the methylation noise on all receptor clusters, $\langle\eta_m^2 \rangle= 2 \RT p(1-p)/(\alpha \tm)$, where we have assumed that the methylation noise on the respective receptor clusters is independent. The phosphorylation noise has strength $\langle\eta_x^2 \rangle = 2 \xT f(1-f)/\tr$.

\subsection{Parameter values}
A large body of work has studied the parameters of the MWC model for the \emph{E. coli} chemotaxis system. We have listed the parameters relevant for our model in table \ref{tab:micropars}. We choose the background concentration $\bar \ell$ to be in between $\KDI$ and $\KDA$, at $\bar \ell =100\mu \mathrm{M}$. 

\begin{table}[b]
	\caption{\label{tab:micropars} \textbf{Measured \textit{E. coli} chemotaxis parameter values.}}
	\begin{ruledtabular}
		\begin{tabular}{llll}
			\textit{Parameter} & \textit{Value} & \textit{Source} & \textit{Description}\\
			\colrule
		$\KDI$ & $18\mathrm{\mu  M}$ &\text{\cite{sourjik_functional_2004,mello_effects_2007}} & MeAsp-Tar dissociation constant inactive receptor \\
		$\KDA$ & $2900\mathrm{\mu  M}$ & \text{\cite{sourjik_functional_2004,mello_effects_2007}} & MeAsp-Tar dissociation constant active receptor \\
		$N$ & $\sim 6$ & \text{\cite{shimizu_modular_2010, sourjik_functional_2004, mello_effects_2007, levit_organization_2002}} & Number of receptors per cluster\\
		$\tilde \alpha$ & $2 k_{\rm B} T$ &\text{\cite{shimizu_modular_2010}} & Free energy change per added methyl group \\ 
		$p$ & $\frac{1}{3}$, $\frac{1}{2}$ &\text{\cite{shimizu_modular_2010, mello_effects_2007} }& Steady state activity at $22\degree C$, $32\degree C$ \\
		$\tr$ & $\sim0.1s$ & \text{\cite{govern_optimal_2014,mattingly_escherichia_2021, levit_organization_2002}} & Phosphorylation timescale \\
		\end{tabular}
	\end{ruledtabular}
\end{table}

In this work we analyze the impact of the methylation timescale $\tm$, and the numbers of receptor clusters and readout molecules $\RT$ and $\xT$, on the past and predictive information. We therefore do not set them to a fixed value, but experimental estimates are listed in table \ref{tab:varpars}. 

\begin{table}[tbh]
	\centering
	\caption{\label{tab:varpars} \textbf{Approximate \emph{E. coli} chemotaxis timescales and abundances.}}
	\begin{ruledtabular}
		\begin{tabular}{llll}
			\textit{Parameter} & \textit{Value} & \textit{Source} & \textit{Description}\\
			\colrule
			$\tm$ & $\sim 10s$ &\text{\cite{segall_temporal_1986, shimizu_modular_2010, mattingly_escherichia_2021}} & Adaptation time \\
		Tsr+Tar & $14000, 3300$ &\text{\cite{li_cellular_2004}} & Rich medium; RP437, OW1 strain \\
		Tsr+Tar & $24000, 37000$ &\text{\cite{li_cellular_2004}} & Minimal medium; RP437, OW1 strain \\
		CheY & $8200, 1400$ & \text{\cite{li_cellular_2004}} & Rich medium; RP437, OW1 strain\\
		CheY & $6300, 14000$ &\text{\cite{li_cellular_2004}} & Minimal medium; RP437, OW1 strain\\
		\end{tabular}
	\end{ruledtabular}
\end{table}

\subsection{Model statistics}
\label{sec:CN:stats}
Again we take the power spectrum route to determine the variance in
the network output, the SNR, and the correlation coefficient between current
output and the future signal. We consider the system to sense the
non-Markovian ligand concentration defined in equation
\eref{ho}. Such a signal is characterized by both its
concentration and derivative, and the (cross-)power spectra of these
properties are
\begin{align}
	\mathcal{S}_s(\omega)= \begin{pmatrix}
		S_\ell(\omega) & S_{\ell\to v}(\omega) \\
		S_{v\to \ell}(\omega) & S_v(\omega) \end{pmatrix}&= \begin{pmatrix}
		S_\ell(\omega) & i\omega S_\ell(\omega) \\
		-i\omega S_\ell(\omega) & \omega^2 S_\ell(\omega)
	\end{pmatrix},
\end{align}
with
\begin{align}
	S_\ell(\omega) &= \frac{2\sigma_v^2/\tau_v}{(\omega^2+((2\tau_v)^{-1} +\rho)^2)(\omega^2+((2\tau_v)^{-1} -\rho)^2)},
\end{align}
where $\rho = \sqrt{(4\tau_v^2)^{-1}-\omega_0^2}$. The chemotaxis signalling network is fully determined by the following matrices (\eref{signallingOU})
\begin{align}
	\mathcal{G}&= q\begin{pmatrix} 1/(\alpha \tm) & 0 \\ -\gamma & 0 \end{pmatrix}, \\[4ex]
	\mathcal{J} &= \begin{pmatrix} -1/\tm & 0 \\ \alpha\gamma & -1/\tr \end{pmatrix},\\[4ex]
	\mathcal{B} &= \begin{pmatrix}
		\sqrt{\langle\eta_m^2 \rangle} & 0 \\ 0 &\sqrt{\langle\eta_x^2 \rangle}
	\end{pmatrix}.
\end{align}

The Fourier transform of the matrix exponential of the Jacobian is 
\begin{equation}
	\begin{split}
		\mathcal{F}\{e^{\mathcal{J} t}\} &= (i\omega \mathbb{I}_n -\mathcal J)^{-1} \\
		&= \begin{pmatrix}
			\frac{1}{1/\tm+i\omega} & 0 \\[2ex]
			\frac{\alpha \gamma}{(1/\tm+i\omega)(1/\tr+i\omega)} & \frac{1}{1/\tr+i\omega}
		\end{pmatrix},
	\end{split}
\end{equation}
which allows us to determine the gain matrix via $\mathbb{G}(\omega) =  \mathcal{F}\{e^{\mathcal J t}\}(\omega)\mathcal{G}$, and the noise matrix using $\mathbb{N}(\omega) =  \mathcal{F}\{e^{\mathcal J t}\}(\omega)\mathcal{B}$;  see also \eref{fouriersol} and \eref{powspecgeneral}.

To gain more insight in the way in which the network maps the signal onto its output, we first study the integration kernels of the system. The integration kernel from ligand concentration to output is given by the inverse Fourier transform of element $(1,2)$ of the gain matrix $\mathbb{G}(\omega)$, which is
\begin{align}
	k(t) &\equiv \mathcal{F}^{-1}\{\tilde g_{\ell \to x}(\omega)\} = \kappa N f(1-f)(1-p) \xT\frac{1}{1-\tr/\tm} \left(\frac{1}{\tm}e^{-\tau/\tm} - \frac{1}{\tr}e^{-\tau/\tr}\right),
\end{align}
with $\kappa = (\bar \ell +\KDI)^{-1}-(\bar \ell
+\KDA)^{-1}$. Due to the adaptive nature of the network, the static
gain from ligand concentration to output is zero: 
$\bar g_{\ell \to x} = \int^\infty_0 k(t)dt = 0$; the long-time
response to a step change in a constant input is zero.  The kernel
does indeed not change the output based on the input
concentration directly, but instead takes a (time-averaged)
derivative of the input (\fref{kernspec}A). It is
therefore useful to consider the kernel that maps the signal
derivative onto the output. This kernel can be found by rearranging
the expression for the output of a linear signalling network,
\eref{dx}. Disregarding the noise terms and integrating by parts
gives
\begin{equation}
	\int_{-\infty}^0 k(-t) \ell(t)dt = K(-t)\ell(t)|^0_{-\infty} - \int_{-\infty}^0 K(-t) v(t)dt,
\end{equation}
where $v(t)\equiv\dot \ell$ and $K(t)$ is the primitive of $k(t)$. To make progress we first determine $K(t)$,
\begin{equation}
	K(t) = \kappa N f(1-f)(1-p) X_T \frac{1}{1-\tr/\tm}\left(-e^{-\tau/\tm} +e^{-\tau/\tr}\right).
\end{equation}
The form of $K(t)$ is that of a simple exponential kernel with a delay (\fref{kernspec}B). We thus have both $K(0) = 0$ and $K(\infty)=0$. It is now clear that the convolution over the ligand concentration simply maps onto the convolution over its derivative as
\begin{equation}
	\int_{-\infty}^0 k(-t) \ell(t)dt =  - \int_{-\infty}^0 K(-t) v(t)dt.
\end{equation}
The static gain of $K(t)$ is $\bar g_{v \to x} = \int_0^\infty K(t) dt =q\gamma\tr\tm=\kappa N \xT (1-p)f(1-f)\tm$. The gain thus increases with the number of receptors per cluster, $N$, the number of readout molecules, $\xT$, and notably, with the adaptation time $\tm$. This static gain from signal derivative to network output is a useful quantity which we will use to describe the other statistics of the network below.

\begin{figure}[tbhp]
	\centering
	\includegraphics[width=\linewidth]{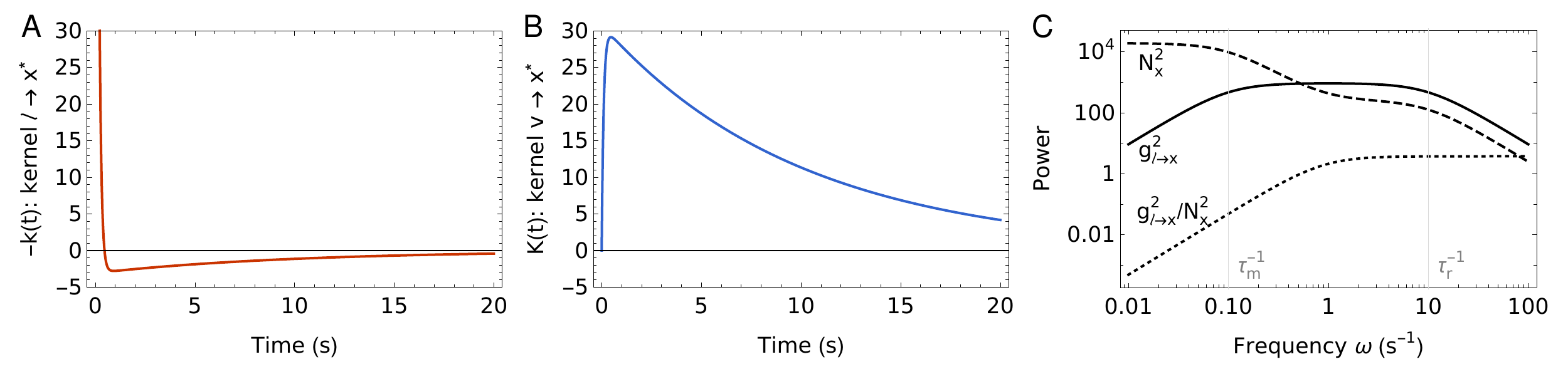}
	\caption{\textbf{Integration kernel and power spectra.} (A) The integration kernel $k(t)$ takes a temporal derivative by weighing the most recent signal values with an opposite sign from the preceding ones. (B) The integration kernel $K(t)$ from the derivative of the input concentration to the network output. The kernel $K(t)$ is the primitive of $k(t)$, and its static gain is proportional to the adaptation timescale $\tm$. (C) Frequency dependent gain $\tilde g_{\ell \to x}^2(\omega)$, frequency dependent noise $N^2_x(\omega)$, and their ratio, as a function of frequency. The chemotaxis network is a band-pass filter, the frequencies that are passed through are set by $\tr$ on the high end and $\tm$ on the low end. At low frequencies, the methylation noise dominates. Parameters used in all panels $\tr=0.1\mathrm{s}$ and $\tm=10\mathrm{s}$. Model parameters are $\tilde a =2$, $N=6$, $\KDI=18 \mathrm{\mu M}$, $\KDA=2900 \mathrm{\mu  M}$, $\bar \ell = 100 \mathrm{\mu  M}$, $p = f =0.5$.}
	\label{fig:kernspec}
\end{figure}

To compute the past and predictive information, we need to determine the variance in the output, the SNR, and the correlation between the current output and the future ligand derivative. To that end we require the power spectrum of the output, and the cross-spectrum from output to future derivative. For the power spectrum of the output we use \eref{powspecgeneral} to find
\begin{equation}
	\elabel{powspecchemx}
	S_x(\omega) = \frac{ q^2\gamma^2\omega^2}{(\tr^{-2}+\omega^2)(\tm^{-2}+\omega^2)} S_\ell(\omega) + \frac{\alpha^2\gamma^2 \langle \eta_m^2\rangle}{(\tr^{-2}+\omega^2)(\tm^{-2}+\omega^2)} + \frac{\langle \eta_m^2\rangle}{\tr^{-2}+\omega^2}.
\end{equation}
From this power spectrum we can see that the network is a band-pass filter, where the gain is maximal in the frequency range
$\tm^{-1}<\omega<\tr^{-1}$. Both for $\omega \gg \tr^{-1}$ and $\omega \ll \tm^{-1}$ the gain goes to $0$. On long timescales the methylation noise dominates (\fref{kernspec}C). The cross-power spectrum between current output and future ligand derivative is given by element $(2,2)$  of the matrix $\mathbb{G}(-\omega) \mathcal{S}_s(\omega)$ which is (also see \eref{crosspowspec})
\begin{equation}
	\begin{split}
		S_{x\to v}(\omega)=  q\gamma\frac{-\omega^2  S_\ell(\omega)}{(\tm^{-1}-i\omega)(\tr^{-1}-i\omega)}.
	\end{split}
\end{equation}

In the main text, we argue that the biologically relevant regime of the input signal is the limit $\omega_0 \to 0$. We therefore present below the network statistics in this limit. We start by determining the variance in the readout, via the inverse Fourier transform of its power spectrum (\eref{powspecchemx}):
\begin{equation}
	\elabel{varchemx}
	\begin{split}
		\lim_{\omega_0\to 0} \sigma_x^2 &= \bar g_{v \to x}^2\frac{1+\tr/\tv + \tr/\tm}{(1+\tm/\tv)(1+\tr/\tv)(1+\tr/\tm)}\sigma_v^2 + \bar g^2_{a\to x} \alpha \RT p(1-p)\frac{1}{1+\tr/\tm}+ \xT f(1-f),\\
		&=\underbrace{\bar g_{v \to x}^2\frac{1+\tr/\tv + \tr/\tm}{(1+\tm/\tv)(1+\tr/\tv)(1+\tr/\tm)}}_{\text{dynamical gain}}\sigma_v^2 + \xT f(1-f)\left(1+\bar g_{v \to x}\frac{\tilde \alpha (1-p)}{\RT \kappa \tm}\frac{1}{1+\tr/\tm}\right),
	\end{split}
\end{equation}
where $\bar g_{a\to x} = \gamma \tr = \xT f(1-f)/(\RT p)$ is the static gain from receptor activity to readout, and we used the definition of $\alpha = \tilde \alpha N p(1-p)$. Because there is no receptor-ligand binding noise, there is also no time averaging as in the push-pull network (and hence no factor depending on $\tr/\tc$). There is methylation noise on a timescale $\tm$, but this cannot be time-averaged effectively because the integration time $\tr$ of the push-pull network is shorter than the receptor methylation timescale $\tm$. The methylation noise can only be averaged out significantly by increasing $\RT$. The contribution from the variance in the signal derivative, $\sigma^2_v$, to the output noise $\sigma^2_x$, depends on the dynamical gain, which is the product of the static gain $\bar{g}^2_{v\to v}$ and a factor that only depends on ratios of timescales. The dynamical gain is maximized for $\tr\to 0$ and $\tm \to \infty$, which is intuitive since subtracting a signal from an earlier one reduces the amplification of the signal.
Hence, when the system has too few $\xT$ molecules to lift the signal above the noise, $\tm$ must be increased to raise the gain. Only when $\xT$ is sufficiently large, can $\tm$ be reduced. This allows the system to take more recent derivatives. The signal to noise ratio $\text{SNR}=\sigma_{x|\eta}^2/ \sigma_{x|L}^2$ can straightforwardly be obtained from \eref{varchemx}. For the covariance between the current output and the future derivative we have
\begin{equation}
	\elabel{futcorcn}
	\begin{split}
		\lim_{\omega_0\to 0} \langle \delta x(0)\delta v(\tau)\rangle &= \mathcal{F}^{-1}\{S_{x\to v}(\omega)\},\\
		&= \frac{-\bar g_{v \to x} \sigma_v^2 }{ (1+\tm/\tau_v)(1+\tr/\tau_v)} e^{-\tau/\tau_v}.
	\end{split}
\end{equation}
The variance in \eref{varchemx} can be used to obtain the normalized correlation function $\langle \delta x(0)\delta v(\tau)\rangle/(\sigma_x\sigma_v)$. 

\subsection{Past and predictive information of the chemotaxis network} \label{sec:CN:IpastIpred} The past and predictive information are straightforward to compute from the quantities above. The definition of the past information is the same as for the push-pull network, and is given by \eref{ipastdef1}. The SNR is now given by, using \eref{varchemx}:
\begin{equation*}
	\text{SNR} = \sigma_{x|\eta}^2/ \sigma_{x|L}^2 =\kappa^2 N \tau_m^2 \sigma_v^2  \frac{1+\tr/\tv + \tr/\tm}{(1+\tm/\tv)(1+\tr/\tau_v)(1+\tr/\tm)}\bigg/\left(\frac{1}{N \xT f(1-f)(1-p)^2}+\frac{\tilde \alpha}{\RT(1+\tr/\tm)}\right),
\end{equation*}
where $\kappa= (\bar \ell +\KDI)^{-1}-(\bar \ell +\KDA)^{-1}$. The predictive information is found in the same manner as in \eref{iprednew}, but now it is a function of the correlation between the current output and the future {\em derivative} of the ligand concentration. This correlation can be decomposed into the instantaneous correlation coefficient and an exponential decay on the timescale of the fluctuations of the derivative of the concentration, \eref{futcorcn}. Specifically, the predictive information is given by
\begin{equation}
	\elabel{ipredcn}
	I_\text{pred}(x_0 ;v_\tau) = -\frac{1}{2}\log(1-\rho_{\ell v}^2 e^{-2\tau/\tau_v}) .
\end{equation}
The instantaneous correlation coefficient $\rho_{\ell v}^2$  can be found using \eref{futcorcn} and \eref{varchemx}. From \eref{ipredcn} it is clear that just like for the push-pull network, the optimal design of the network that maximizes the predictive information, determined by the optimal ratio $\xT/\RT$ and the optimal adaptation time $\tm$, does not depend on the forecast interval $\tau$. The forecast interval only affects the magnitude of the predictive information.

\subsection{Optimal allocation}
We can determine the optimal ratio $(\xT/\RT)^\text{opt}$ that maximizes either the past information or the predictive information, given all other network parameters, most notably $\tm$. Just as for the push-pull network, we find however that the optimal ratio $(\xT/\RT)^\text{opt}$ is the same regardless of whether the past or the predictive information is maximized. This is again because the information on the future signal (be it the value or the derivative) is encoded in the receptor occupancy, while the ratio $\xT/\RT$ controls the interval by which the downstream readout samples the receptor to estimates its occupancy. Nonetheless, the optimal methylation timescale $\tm^\text{opt}$ that maximizes either the past or the predictive information is different---maximizing predictive information requires a more recent derivative and hence a shorter $\tm$ than obtaining past information.

Given $\tm$ and all other parameters, the optimal ratio of the number of readout molecules over receptor clusters is, using $C=\RT+\xT$,
\begin{equation}
	\begin{split}
		\left(\frac{\xT}{\RT}\right)^\text{opt} &= \sqrt{\frac{1}{\alpha}\frac{1}{f(1-f)}\frac{p}{1-p}}\sqrt{1+\frac{\tr}{\tm}}, \\
		&=2\sqrt{2/N}\sqrt{1+\frac{\tr}{\tm}},
	\end{split}
\end{equation}
where in the second line we have used that $\alpha = \tilde \alpha N p(1-p)$, and $\tilde \alpha=2$, and $f=p=0.5$. Because for the chemotaxis network $\tr<\tm$ the ratio $\tr/\tm$ only varies between $0$ and $1$. For this reason, the optimal ratio $(\xT/\RT)^\text{opt}$ depends only weakly on $\tm$, and does not vary strongly along the isocost lines of Fig. 4A in the main text, see \fref{XtRtovertheta}.

\begin{figure}[tbhp]
	\centering
	\includegraphics[width=0.4\linewidth]{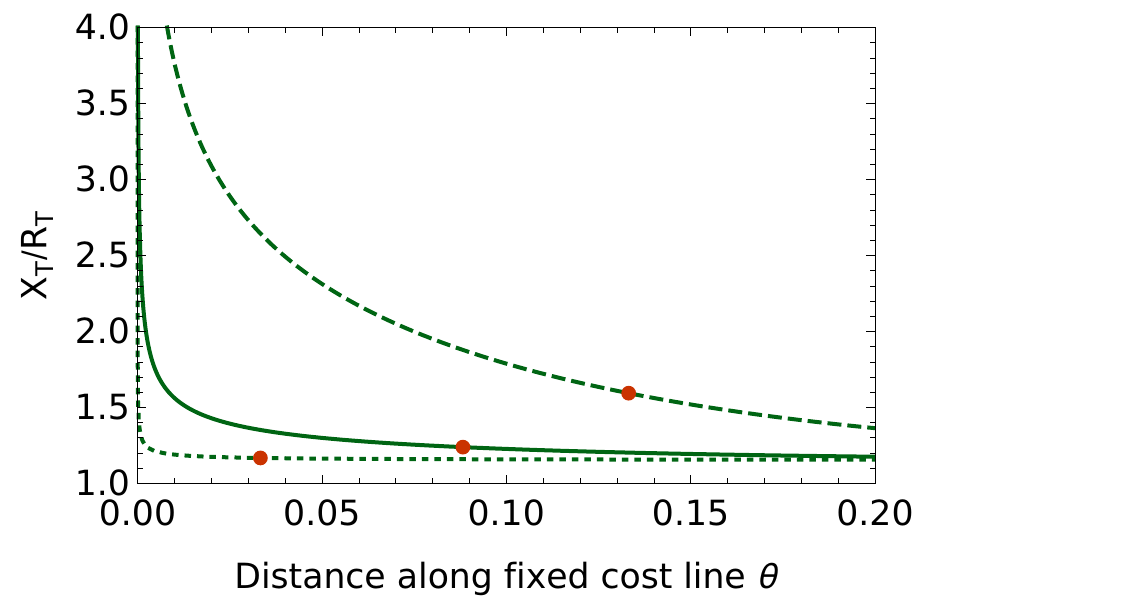}
	\caption{\textbf{The optimal allocation ratio $\xT/\RT$ varies only slightly along the isocost lines of Fig. 4A in the main text.} The optimal ratio $\xT/\RT$ as a function of the distance $\theta$ along the isocost lines of Fig. 4A of the main text; dotted line $C=10^2$, solid line $C=10^4$, dashed line $C=10^6$. The red dots mark the points where the predictive information is maximal. Along the isocost lines $\xT/
		\RT$ varies much more weakly than for the push-pull network; for resource availability $C\leq 10^4$ the ratio is almost constant. Parameters used $g=4 \mathrm{mm}^{-1}$, $\tr=0.1s$, $\KDI=18 \mathrm{\mu M}$, $\KDA=2900 \mathrm{\mu M}$, $N=6$, $\tilde \alpha =2$, $p=f=0.5$, $\bar \ell = 100 \mathrm{\mu M}$.}
	\label{fig:XtRtovertheta}
\end{figure}


\newpage
\twocolumngrid

\bibliography{library_age}

\end{document}